\documentclass[fleqn,usenatbib]{mnras}

\usepackage[T1]{fontenc}

\DeclareRobustCommand{\VAN}[3]{#2}
\let\VANthebibliography\thebibliography
\def\thebibliography{\DeclareRobustCommand{\VAN}[3]{##3}\VANthebibliography}

\usepackage{graphicx}	
\usepackage{amsmath}	
\usepackage{amssymb}	

\usepackage{newtxtext,newtxmath}

\usepackage{placeins}

\title[Map-based simulations of systematics]{Fast map-based simulations of systematics in CMB surveys including effects of the scanning strategy}

\author[N. McCallum et al.]{
Nialh McCallum,$^{1}$\thanks{E-mail: nialh.mccallum@postgrad.manchester.ac.uk}
Daniel. B. Thomas$^{2}$
and Michael L. Brown$^{1}$
\\
$^{1}$Jodrell Bank Centre for Astrophysics, School of Physics \& Astronomy, The University of Manchester, Manchester M13 9PL, UK\\
$^{2}$School of Physics and Astronomy, Queen Mary University of London, London, E1 4NS, UK\\
}

\date{Accepted XXX. Received YYY; in original form ZZZ}

\pubyear{2021}

\begin{document}
\label{firstpage}
\pagerange{\pageref{firstpage}--\pageref{lastpage}}
\maketitle

\begin{abstract}
We present approaches to quickly simulate systematics affecting CMB observations, including the effects of the scanning strategy. Using summary properties of the scan we capture features of full time ordered data (TOD) simulations, allowing maps and power spectra to be generated at much improved speed for a number of systematics -- the cases we present experienced speed ups of 3-4 orders of magnitude when implementing the map-based approaches. We demonstrate the effectiveness of the approaches at capturing the salient features of the scan by directly comparing to full TOD simulations -- seeing agreement at sub-percent levels of accuracy. We simulate the effects of differential gain, pointing, and ellipticity to show the effectiveness of the approaches, but note that one could extend these techniques to other systematics. We finally show how to apply these fast map-based simulations of systematic effects to a full focal plane showing their ability to incorporate thousands of detectors as seen in modern CMB experiments.
\end{abstract}




\section{Introduction}
The cosmic microwave background (CMB) anisotropies provide one of the most powerful observational probes of the standard cosmological model \cite[for recent reviews see e.g.][]{durrer2015, staggs2018}. A key target of many future CMB experiments e.g. LiteBIRD, Simons Observatory (SO), BICEP and Keck Array, CMB-S4, \citep{2018JLTP..193.1048S,2019JCAP...02..056A,2018SPIE10708E..07H,2016arXiv161002743A} is the observation of the low multipole $B$-mode polarization signal related to primordial gravitational waves. $B$-modes also arise on smaller angular scales via gravitational lensing of the $E$-mode signal, which has been observed by a number of experiments e.g. Polarbear, \citep{2020ApJ...893...85F}.

Upcoming CMB experiments are targeting unprecedented levels of sensitivity as they target the low amplitude (potentially vanishingly small) signature from the primordial $B$-mode signal; for example a primary objective of upcoming Stage IV experiments is constraining the tensor-to-scalar ratio to an accuracy of $\Delta r \approx 0.001$ \citep{2016arXiv161002743A}. These ambitious targets will require exquisite understanding and control of foreground contamination \citep{2016arXiv160603606D}, delensing \citep{2017JCAP...12..005G}, and instrumental systematics \citep{2003PhRvD..67d3004H}.

The impact of systematics on CMB polarization measurements has been an important consideration for recent experiments. A plethora of studies of systematic effects have been undertaken by many experiments, including satellite experiments such as WMAP \citep{2011ApJS..192...14J}, {\it Planck} \citep{2014A&A...571A...7P,2016A&A...594A...7P}, {\it EPIC} \citep{2009arXiv0906.1188B}, and CORE \citep{2018JCAP...04..022N} and ground-based telescopes such as BICEP2 \citep{2015ApJ...814..110B}, and SO \citep{2018SPIE10708E..3ZC}. As such a number of tools have been developed to deal with a number of different systematic effects. These tools along with new developments will become even more important for next generation surveys as systematics increasingly become the limiting factor in CMB polarization surveys.

One particularly concerning set of systematics are those that cause signals to mix, resulting in coupling of the much larger intensity signal to polarization and mixing between $E$ and $B$-mode polarization signals. Due to the relative size of the signals these can result in spurious $B$-mode signals which would be detrimental to the search for the primordial gravitational wave signature.

A number of these systematics are closely coupled to the scanning strategy, these have been well studied, \cite[see e.g.][and references therein]{2003PhRvD..67d3004H, 2007MNRAS.376.1767O, PhysRevD.77.083003, 0806.3096,Wallisetal2016}. The literature presents a number of approaches to tackle these systematics such as mitigation through a well designed scanning strategy. In particular a number of symmetries exist which may be exploited to heavily suppress systematics of certain spin which has been pointed out in e.g. \cite{2007MNRAS.376.1767O,PhysRevD.77.083003,0806.3096,2015ApJ...814..110B,2020MNRAS.491.1960T}. Additionally a number of formalisms exist which provide details of how to predict the leakage caused by systematics at the power spectrum level using the full coupling of the scan strategy, systematics, and on sky signals \cite[e.g.][]{2017A&A...598A..25H,2020arXiv200800011M}.

Further mitigation of these types of systematics is possible through the use of a continuously rotating or stepped half-wave plate (HWP) \citep{brown2009, 2018SPIE10708E..48S}. However, systematic effects associated with the HWP itself can also be present and careful assessment of these will be required for future experiments intending to use them \citep{2004MNRAS.349..321B,2019A&A...627A.160D}. For future experiments that do not include a HWP, techniques such as pair differencing may be employed to separate the intensity and polarization signal. In fact even when a HWP is included pair differencing is often assumed during the experiment design in order to quickly model the effect of systematics, e.g. as was done recently for the SO small aperture telescopes, which will include HWPs \citep{2018SPIE10708E..3ZC}. The approaches we explore in this work do not specifically include a HWP and we leave it to future work to include these.

A common way to investigate the effects of systematics on CMB measurements is to run full time ordered data (TOD) simulations -- looping over pointing data while injecting systematics and noise at each step. These are a computationally intensive endeavour and given the constraints of upcoming experiments (e.g. the resolution and number of detectors etc.) are becoming increasingly computationally expensive. There are of course no perfect replacements to these full realistic TOD simulations and in recent years this has led to great improvements in the efficiency of these simulations particularly in the time consuming implementation of beam convolution \cite[see e.g.][]{2011ApJS..193....5M, 2014MNRAS.442.1963W,2019MNRAS.486.5448D,2020JCAP...02..030F}. However these remain computationally heavy processes and as such implementations which can mimic some of the salient features of a full TOD simulation in a less time costly approach are desirable.

We present two approaches to do this. Our first treatment is a map level approach which is an extension of the work of \cite{Wallisetal2016} and \cite{2020arXiv200800011M}, involving the direct multiplication of maps containing averaged scan information with systematic sourced maps. For a number of spin-coupled systematics this performs very well in terms of reproducing the effects seen in a full TOD simulation. Our second technique is a constant elevation scan (CES) approach, building on the work of \cite{Wallisetal2016}, \cite{2020arXiv200800011M} and \cite{2021arXiv210202284T}, which utilises the very significant constraints that CESs place on the allowed crossing angles in a ground-based scanning strategy to shortcut the TOD process. While complex time dependent effects are beyond the scope of these approaches, they nevertheless offer a method to shortcut the TOD process by simulating a number of time-independent systematics directly in map-space while retaining features of the scan.

These approaches will aid in speeding up studies of systematics for upcoming surveys by avoiding the need to repeatedly run time costly TOD simulations. By running a full TOD simulation just once we may then utilise stored maps which incorporate features of the scanning strategy -- these then facilitate the fast map-based approaches which can forecast the effects of a number of systematics with varying magnitudes. The map-based approaches readily extend to full focal plane simulations with the ability to rapidly simulate systematics for thousands of detectors as seen in modern CMB experiments, allowing studies of correlated systematic effects between detectors. They can also be used in tandem with multiple CMB realisations enabling monte-carlo techniques to be implemented at much greater speeds with scanning strategy effects incorporated.

The paper is organised as follows. In section \ref{section:Map-Based Sims} we present a fast map-based approach to simulate systematics while retaining structure of the scanning strategy. We then apply this approach to differential gain and pointing systematics in section \ref{section:Demonstration} showing a significant speed up when comparing to TOD simulations. In section \ref{section:CES Approach} we present a second approach, based on CES constraints, that facilitates rapid simulations of systematic signals which vary with crossing angle -- we demonstrate this approach by applying it to the case of differential ellipticity. Finally we demonstrate how to apply the fast map-based approach to a full focal plane (consisting of a large number of detectors) in section \ref{section:Full Focal Plane}, and show the effects that correlated systematics across detectors can have on the results. In section \ref{section:Conclusion} we summarise our results.

\section{Map-Based Simulations of spin-coupled systematics}
\label{section:Map-Based Sims}
As standard, CMB experiments aim to measure both the spin-0 intensity and the spin-2 polarization.\footnote{The spin dependence we refer to is the dependence of the fields with crossing angle $\psi$, which gives the orientation of the scan direction of the instrument with respect to North, such that a spin-$\pm n$ field $d_n^Q \pm id_n^U$ contributes to a detector timestream as $d_n^Q\cos(n\psi_j)+d_n^U\sin(n\psi_j)$.} The way in which these signals are disentangled from one another will depend on a number of factors which are somewhat experiment specific -- the presence of a HWP, redundancies of a scanning strategy, or the use of pair differencing amongst other considerations will dictate how analysis choices for the data are eventually made.

An individual polarized detector will measure the on-sky signal according to the detector equation
\begin{equation}
    d = I + Q\cos(2\psi) + U\sin(2\psi)
    \label{equation:Detector Equation}
\end{equation}
where $\psi$ is the crossing angle, and $I$, $Q$ and $U$ are the Stokes parameters. In addition there will be contamination by systematics and noise.

In order to measure the intensity and polarization signals one could employ the familiar simple binning approach to map-making using \cite[e.g.][]{brown2009}
\begin{equation}
\begin{pmatrix}\hat{I}\\\hat{Q}\\\hat{U}\end{pmatrix}
=
M^{-1}
\begin{pmatrix}\langle d_{j} \rangle\\\langle d_{j}\cos(2\psi_{j}) \rangle\\\langle d_{j}\sin(2\psi_{j}) \rangle\end{pmatrix} \text{,}
\label{eq:3x3 map making}
\end{equation}
where
\begin{equation}
    M = \begin{pmatrix}1&\langle \cos(2\psi_{j}) \rangle&\langle \sin(2\psi_{j}) \rangle\\\langle \cos(2\psi_{j}) \rangle&\langle \cos^{2}(2\psi_{j}) \rangle&\langle \cos(2\psi_{j})\sin(2\psi_{j}) \rangle\\\langle \sin(2\psi_{j}) \rangle&\langle \sin(2\psi_{j})\cos(2\psi_{j}) \rangle&\langle \sin^{2}(2\psi_{j}) \rangle\end{pmatrix}.
\end{equation}
The angle brackets $\langle \rangle$ denote an average over the $j$ measurements in a sky pixel, each of which has an associated angle $\psi_j$, and the hat on the Stokes $\hat{I}, \hat{Q}, \hat{U}$ denotes it is the estimated quantity. One could also trivially extend this to solve for signals of other spin as is discussed in \cite{2020arXiv200800011M}. We note that the results of this work could also be applied to intensity mapping surveys, by considering leakage from systematics into the spin-0 intensity field, as is done in \cite{2021arXiv210708058M}.

One can equivalently write equation \ref{eq:3x3 map making} as
\begin{equation}
\begin{pmatrix}\hat{I}\\\hat{Q}-i\hat{U}\\\hat{Q}+i\hat{U}\end{pmatrix}
=
\begin{pmatrix}1& \frac{1}{2} \tilde{h}_{2} & \frac{1}{2} \tilde{h}_{-2} \\ \frac{1}{2} \tilde{h}_{2} & \frac{1}{4} \tilde{h}_{4} & \frac{1}{4} \\ \frac{1}{2} \tilde{h}_{-2} & \frac{1}{4} & \frac{1}{4} \tilde{h}_{-4} \end{pmatrix}^{-1}
\begin{pmatrix}\langle d_j \rangle\\\frac{1}{2}\langle d_je^{2i\psi_j} \rangle\\\frac{1}{2}\langle d_je^{-2i\psi_j} \rangle\end{pmatrix} \text{,}
\label{eq:3x3 hn map making}
\end{equation}
where we adopt the $\tilde{h}_n$ quantities used in \cite{Wallisetal2016,2020arXiv200800011M}. This is also similar to the map-making setup presented in \cite{PhysRevD.77.083003}.

The $\tilde{h}_n$ are constructed using knowledge of the crossing angles in the scanning strategy giving the orientation function
\begin{equation}
\begin{split}
    &\tilde{h}_{n}(\Omega) = \frac{1}{N_{\text{hits}}(\Omega)} \sum_{j} e^{in\psi_j(\Omega)}
    \\
    &= \frac{1}{N_{\text{hits}}(\Omega)} \sum_j \left(\cos \left( n\psi_j(\Omega)\right) +i\sin \left( n\psi_j(\Omega)\right) \right)\text{,}
    \label{eq:orientation function}
\end{split}
\end{equation}
which is constructed from the scan by averaging within each pixel. The $\Omega = (\theta,\phi)$ are the latitude and longitude coordinates of each pixel on the sky, and $\psi$ is the orientation angle of the scan direction of the instrument with respect to North. This equation was introduced by \cite{2009arXiv0906.1188B} who pointed out its use for scan strategy design --- a small $\tilde{h}_n$ implies a survey with good crossing angle coverage which aids in suppression of certain systematics. In general, one generates these $\tilde{h}_n$ maps by performing a full TOD simulation once, and storing maps of this scan information (in the same way as the hit map).

The orientation function can be used to write out the coupling of signals of arbitrary spin to a particular signal of interest when performing a scan as
\begin{equation}
    {}_{k}\tilde{S}^{d}(\Omega) = \sum_{k'=-\infty}^{\infty}\tilde{h}_{k-k'}(\Omega){}_{k'}\tilde{S}(\Omega) \text{,}
    \label{eq:spinsyst}
\end{equation}
where $n=k-k'$ shows the term is coupling some signal of spin-$k'$ to the signal of interest of spin-$k$. This relies on the systematics being time independent --- i.e. ${}_{k'}\tilde{S}(\Omega)$ not to vary with time --- over the extent of the data on which the $\tilde{h}_{k-k'}$ are calculated.\footnote{One could in principle include time varying systematics by splitting the TOD of a survey into several chunks of time -- in which the systematic is stable over the chunk but varies between them -- and generating a $\tilde{h}_{k-k'}$ for each of them.} The ${}_{k'}\tilde{S}(\Omega)$ are usually a function of the on-sky signal either directly observed or scaled by some instrumental systematic.

The quantities of equation \ref{eq:spinsyst} are related to the terms appearing on the RHS of the map-making equation (equation \ref{eq:3x3 map making}) as
\begin{equation}
\begin{split}
&\left \langle d_j \right \rangle = {}_0\tilde{S}^{d}\\
&\left \langle d_j \cos 2\psi_j \right \rangle =\Re{}({}_2\tilde{S}^{d})\\
&\left \langle d_j \sin 2\psi_j \right \rangle =\Im{}({}_2\tilde{S}^{d})
\label{eq:RHS Simplification}
\end{split}    
\end{equation}
under the condition that the signals making up $d_j$ have well defined spin dependence i.e. the observed quantity is contributed to by spin-$n$ signals as $d_j = \sum_{n\geq0}(d_n^Q\cos(n\psi_j)+d_n^U\sin(n\psi_j))$ such that the signals $d_n^Q$ and $d_n^U$ do not depend on the crossing angle $\psi_j$ or time. This is the case for many systematics when considering a circular beam as the leaked signals $I$, $Q$, $U$ etc. do not depend on the crossing angle (there would be some slight variation in the signals due to sampling different parts of the pixel but this averages out). This allows us to use the averaged crossing angle quantities ($\tilde{h}_n$) in each pixel to perform one calculation to simulate the effect of systematics with the effects of the scan included, rather than requiring a full TOD which calculates each point in time separately. However in some cases systematics, such as differential ellipticity, lead to a $\psi_j$ dependence of the systematic-induced signal itself which requires a more complicated approach (see section \ref{section:EllipTOD}).

Given equation \ref{eq:RHS Simplification} the map-making equation in the case of time-independent systematics can thus be written as
\begin{equation}
\begin{pmatrix}\hat{I}\\\hat{Q}-i\hat{U}\\\hat{Q}+i\hat{U}\end{pmatrix}
=
\begin{pmatrix}1& \frac{1}{2} \tilde{h}_{2} & \frac{1}{2} \tilde{h}_{-2} \\ \frac{1}{2} \tilde{h}_{2} & \frac{1}{4} \tilde{h}_{4} & \frac{1}{4} \\ \frac{1}{2} \tilde{h}_{-2} & \frac{1}{4} & \frac{1}{4} \tilde{h}_{-4} \end{pmatrix}^{-1}
\begin{pmatrix} \sum_{k'=-\infty}^{\infty}\tilde{h}_{0-k'} {}_{k'}\tilde{S} \\\frac{1}{2}\sum_{k'=-\infty}^{\infty}\tilde{h}_{2-k'} {}_{k'}\tilde{S}\\\frac{1}{2}\sum_{k'=-\infty}^{\infty}\tilde{h}_{-2-k'} {}_{k'}\tilde{S}\end{pmatrix} \text{.}
\label{eq:3x3 hn map making final}
\end{equation}
As such, having stored the relevant $\tilde{h}_{n}$ from a given scan strategy for the systematics of interest one can apply this entire calculation in map-space i.e. using the averaged scan information means all of the quantities involved in equation \ref{eq:3x3 hn map making final} are maps, which facilitates the simple map-based simulation approach of taking the $\tilde{h}_n$ scan maps and multiplying the ${}_{k'}\tilde{S}$ functions of the on-sky signal in map space. The stored $\tilde{h}_{n}$ retain the structure of the scan -- this allows for fast map-based simulations including effects of the scanning strategy to be performed without the need for further full TOD simulations.

\begin{figure*}
  \centering
  \includegraphics[width=0.6666666666\columnwidth]{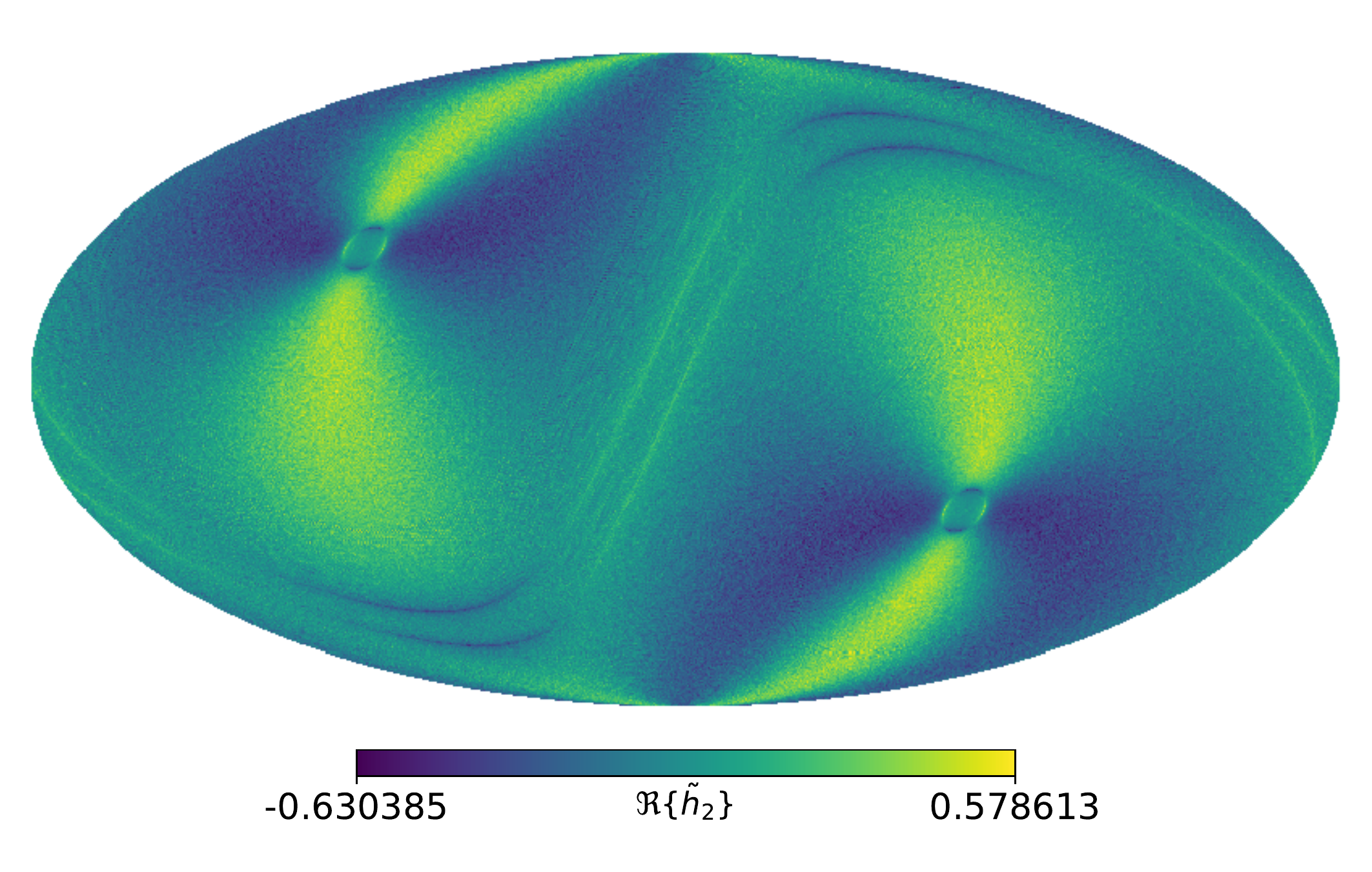}
  \includegraphics[width=0.6666666666\columnwidth]{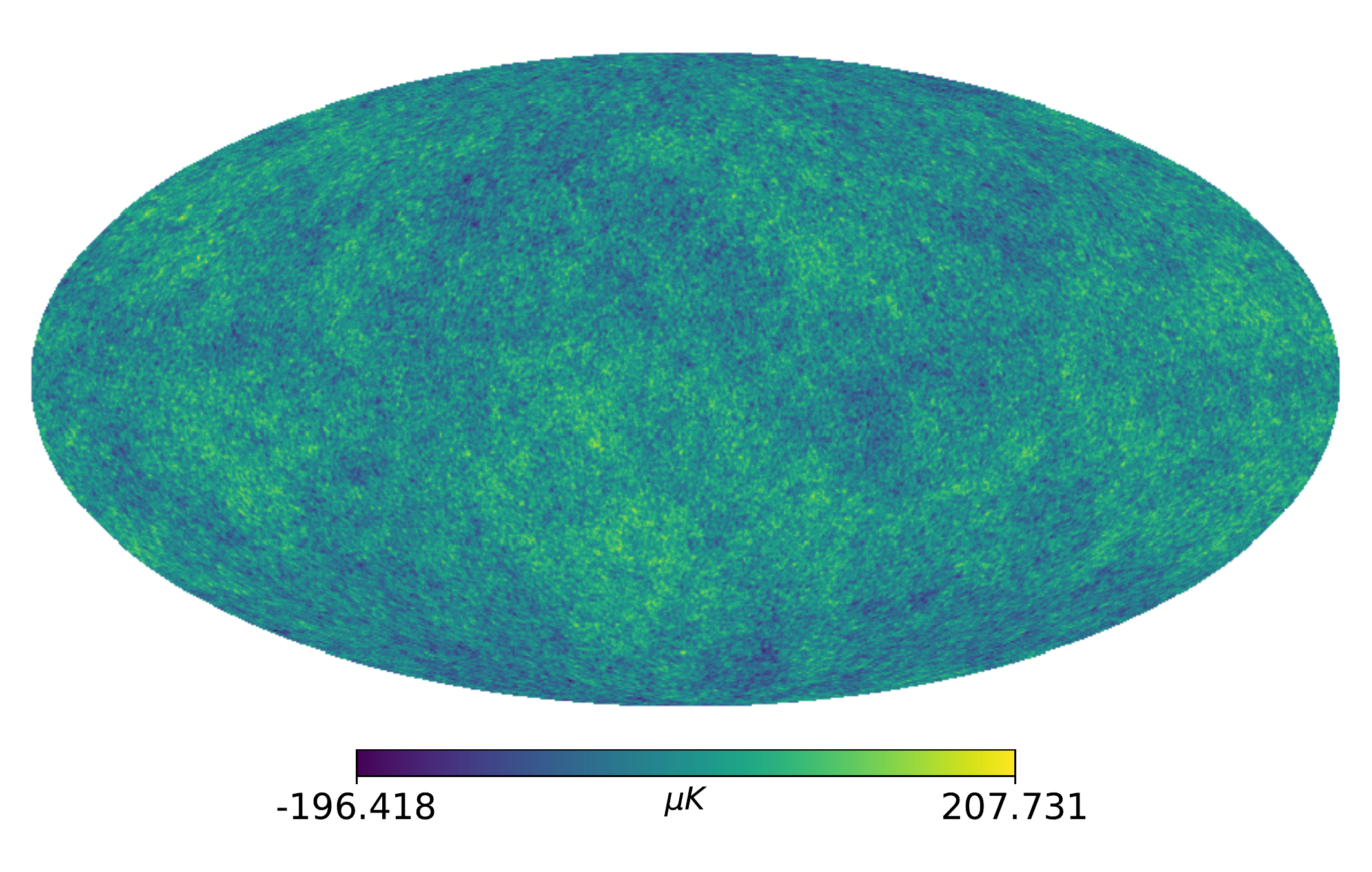}
  \includegraphics[width=0.6666666666\columnwidth]{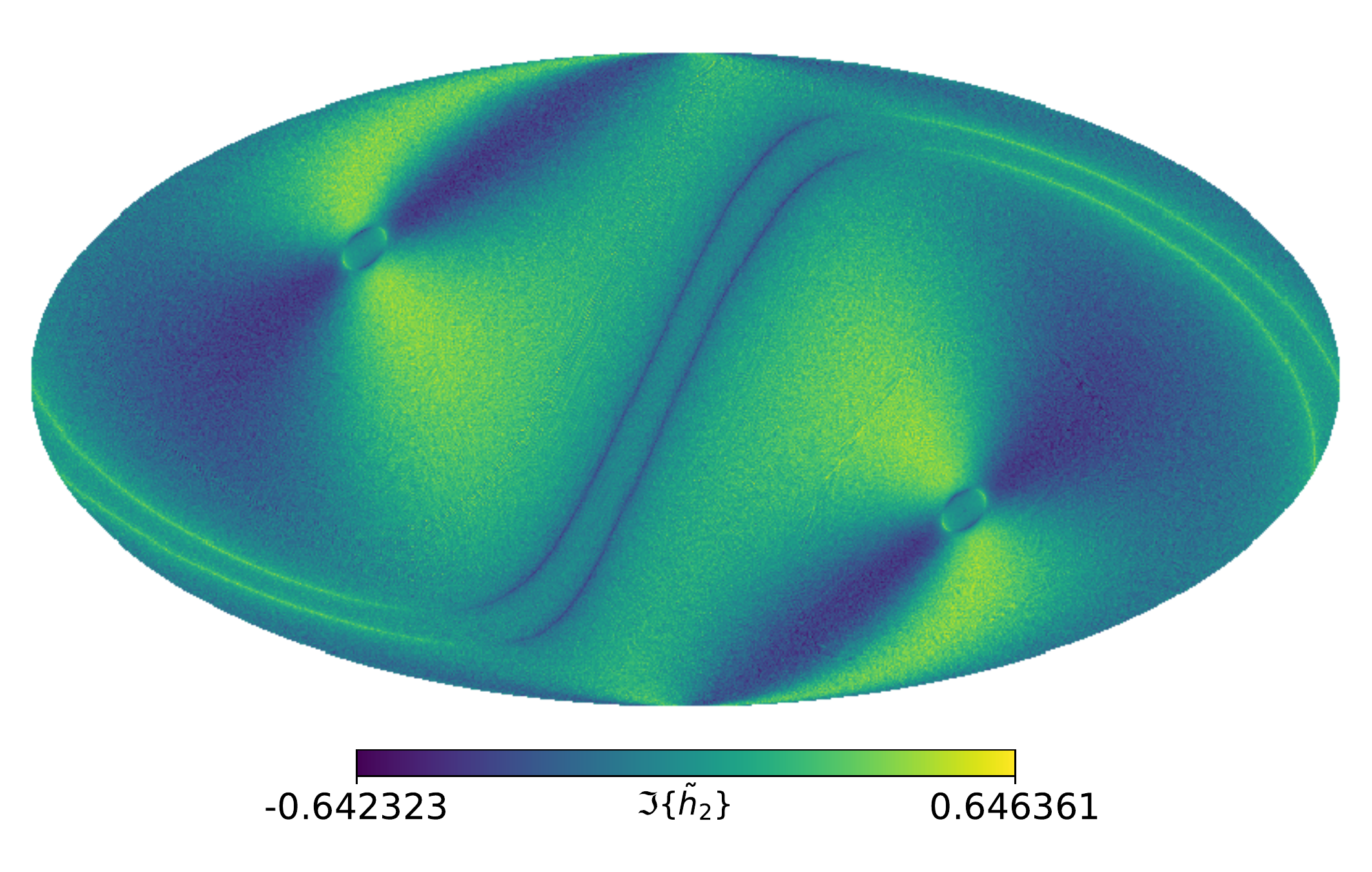}
  
  \includegraphics[width=0.6666666666\columnwidth]{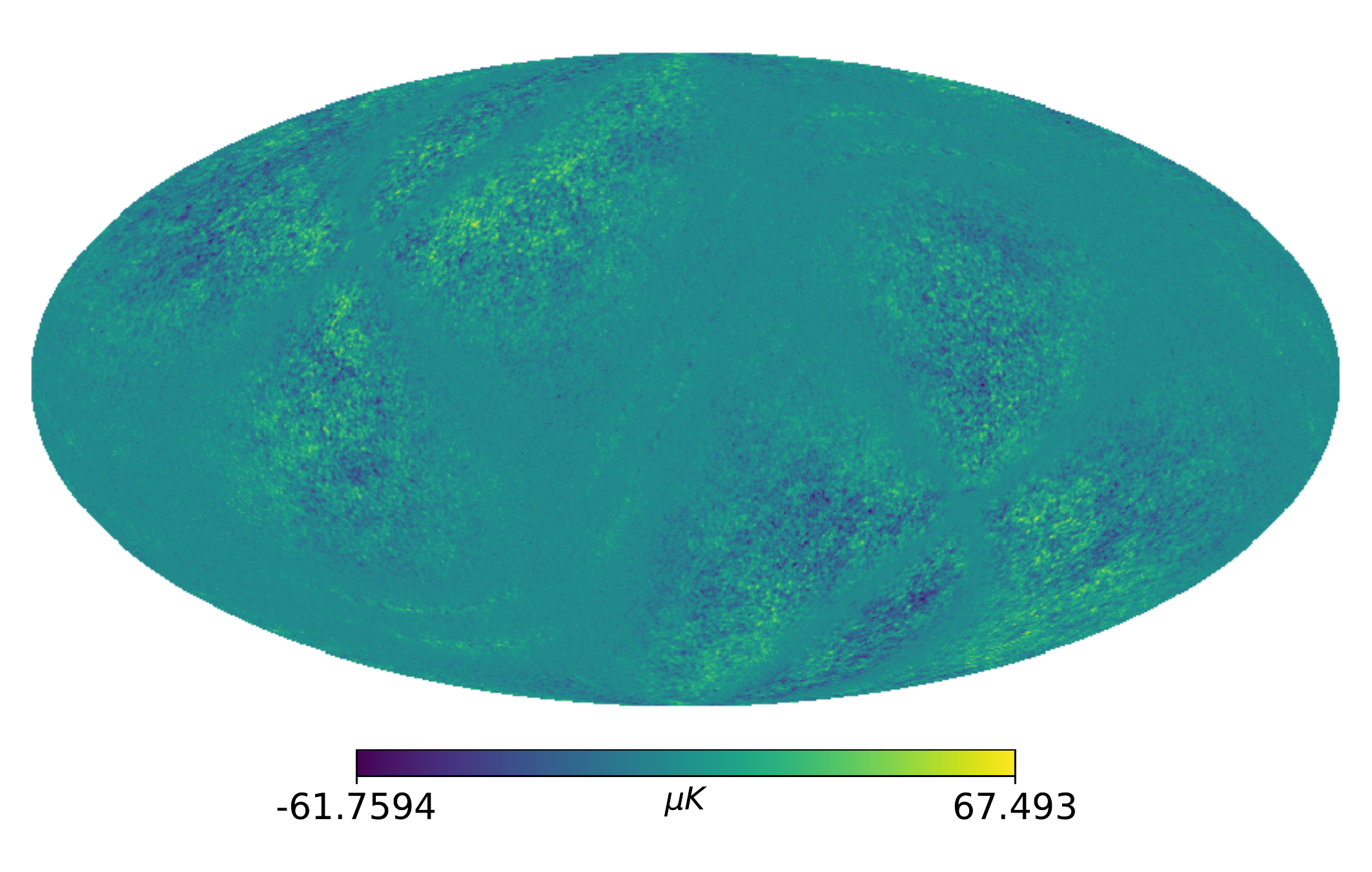}
  \includegraphics[width=0.6666666666\columnwidth]{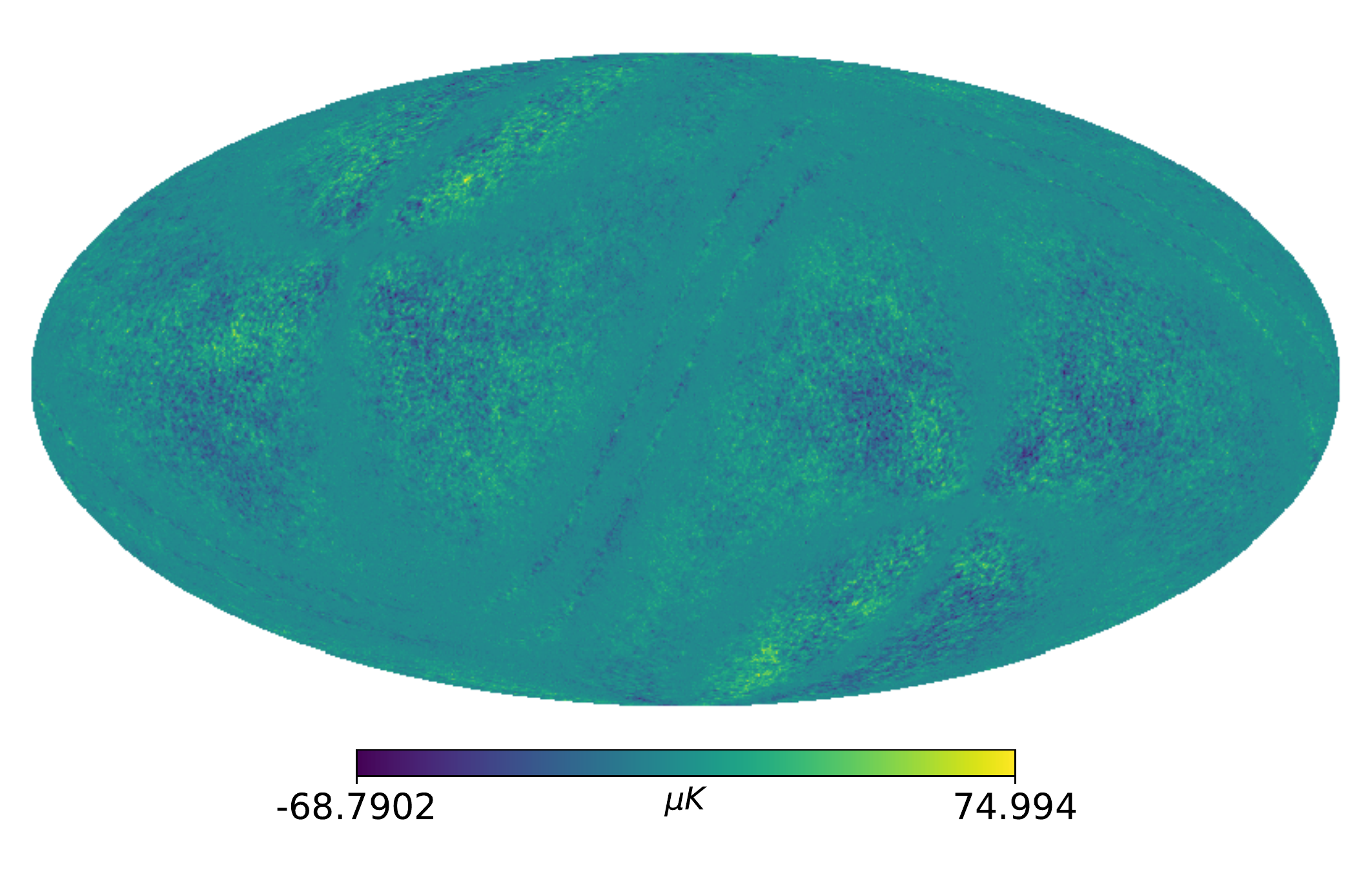}
\caption{Example of the map-based mechanism for spin-0 leakage into spin-2 polarization. The top left panel shows the real part of $\tilde{h}_2$ i.e. $\langle \cos(2\psi_j) \rangle$ which when multiplied by the spin-0 temperature field (top middle panel) gives the leakage field $(\langle d_j \cos(2\psi_j) \rangle)$ (bottom left panel) -- note the clear evidence of tracks in the resultant field which relate to the $\langle \cos(2\psi_j) \rangle$ structure from the {\it EPIC} satellite scanning strategy. Similarly the top right panel shows the imaginary part of $\tilde{h}_2$ i.e. $\langle \sin(2\psi_j) \rangle$ which when multiplied by the spin-0 temperature field (top middle panel) gives the leakage field $(\langle d_j \sin(2\psi_j) \rangle)$ (bottom right panel) -- note again the clear evidence of tracks in the resultant field which relate to the $\langle \sin(2\psi_j) \rangle$ structure from the {\it EPIC} scan.}
\label{figure:Mechanism}
\end{figure*}

In figure \ref{figure:Mechanism} we show an example of how this map-based approach works for the leakage of a spin-0 field into a spin-2 field. Equation \ref{eq:spinsyst} gives this simply as
\begin{equation}
    {}_{2}\tilde{S}^{d}(\Omega) = \tilde{h}_{2}(\Omega){}_{0}\tilde{S}(\Omega) \text{.}
\end{equation}
The top left and right panels of the figure show the real and imaginary parts of the $\tilde{h}_{2}$ quantity for the scanning strategy of the proposed {\it EPIC} satellite \citep{2008arXiv0805.4207B}, and the top centre panel shows the spin-0 temperature field. The resultant leaked signals of $\Re{}({}_2\tilde{S}^{d})$ and $\Im{}({}_2\tilde{S}^{d})$ are shown in the bottom left and right panel respectively -- we see here that the temperature field is present along with track like structures emanating from structure in the $\tilde{h}_{2}$ fields resulting from the scanning strategy. We will show in section \ref{section:Demonstration} that the fast map-based method using multiplication in map space retains scanning structure similar to that in a full TOD simulation.

One can calculate $\tilde{h}_n$ maps simply from the pointing information of a scanning strategy and store the information. By working with these averaged quantities directly the effect of systematics with the scan strategy encoded can be rapidly simulated, without the need to rerun the computationally expensive TOD simulations many times. Furthermore the ${}_{k'}\tilde{S}(\Omega)$ signals are usually some function of the on sky signal e.g. $I,\eth I, Q+iU$ etc. either observed directly or leaked by systematics. As such utilising the speed up of the map-based simulations would facilitate quick monte-carlo simulation techniques with different CMB realisations.

\subsection{Focal Plane Elements}
\label{section:FullFoc1}
The quantities that appear in map-making will depend on how the data is treated by a given experiment. It is possible to use a number of different timestreams during map-making, made from various focal plane elements, common examples include:
\begin{itemize}
    \item a single detector timestream necessitating map-making for temperature and polarization simultaneously -- requring a sufficiently redundant scan strategy.
    \item a single pair of detectors that has been differenced (summed) to solve for polarization (temperature).
    \item a combined timestream from two detector pairs, oriented at 45 degrees to one another in a ``$+\times$'' layout (see section \ref{section:+XSimplification} for further details), allowing simultaneous measurement of Q and U.
\end{itemize}
The methods we use can treat all these cases, but we should note that for a given scanning strategy each would produce different $\tilde{h}_{n}$ terms to include in the map-making equation -- though the scan strategy is the same, the element specific setup such as which detector angles are included will change the associated $\tilde{h}_{n}$. Additionally the systematic leakage affecting each case would differ. The approach we present has been applied to simple binned map-making. In principle, the method could be extended to other map-making approaches (e.g. maximum-likelihood) but the map-making equations (e.g. equation \ref{eq:3x3 map making}) would have a more complicated form than is presented here. We also note that subtleties arising from considering multiple frequencies simultaneously may also complicate the setup.

In section \ref{section:Full Focal Plane} we will build on this focal plane element setup to show a simple way to extend the map-based simulation techniques to a full focal plane. This is achieved by taking the maps made by each focal plane element (e.g. a single detector, a single pair, or single set of ``$+\times$'' pairs) and then averaging them across the full focal plane. An advantage of this is it will translate the speed up afforded by the map-based approach to a full focal plane allowing systematics to be simulated quickly for realistic numbers (thousands) of detectors. This will enable studies of systematics to be performed rapidly for a full focal plane, aiding in forecasting for future CMB experiments without requiring repeated TOD simulations.

\section{Demonstration of the Map-Based simulation technique}
\label{section:Demonstration}
In this section we shall demonstrate the ability of the fast map-based method to mimic a full TOD simulation including the effects of time-independent systematics and the scanning strategy. In order to perform this demonstration we compare the output to a simple full TOD simulation.

\subsection{Detector Setup}
\label{section:+XSimplification}
The detector setup we employ is as in \cite{Wallisetal2016}, where for simplicity two orthogonal detector pairs are considered oriented at $45^{\circ}$ with respect to each other, i.e. a ``$Q$''-like $+$ pair and ``$U$''-like $\times$ pair. This allows for the simultaneous measurement of $Q$ and $U$ polarization. From each detector pair, we use the differenced signal
\begin{equation}
    d = \frac{1}{2} (d^A -d^B),
    \label{eq:differencing}
\end{equation}
where $A$ and $B$ denote the individual detectors within a pair.

An important point to make is that we are including the differenced signal from a TOD element that includes both a $+$ and $\times$ detector pair, each of which will have distinct $\tilde{h}_n(\Omega)$ and signals ${}_{k'}\tilde{S}(\Omega)$. Having different positioning on the focal plane results in altered crossing angle coverage and thus different $\tilde{h}_n(\Omega)$; the positioning on the focal plane can also lead to different ${}_{k'}\tilde{S}(\Omega)$ signals stemming from the on-sky signal as each detector pair observes a slightly different point on the sky. Additionally, the two detector pairs experiencing different levels of systematics can also lead to the ${}_{k'}\tilde{S}(\Omega)$ differing. In this case, equation \ref{eq:spinsyst} must therefore include information for both pairs to construct the total signal as
\begin{equation}
    {}_{k}\tilde{S}^{d,\text{tot}}(\Omega) = \sum_{k'=-\infty}^{\infty} \frac{N^{+}_{\text{hits}}}{N^{\text{tot}}_{\text{hits}}} \tilde{h}^{+}_{k-k'}(\Omega){}_{k'}\tilde{S}^{+}(\Omega) + \frac{N^{\times}_{\text{hits}}}{N^{\text{tot}}_{\text{hits}}}\tilde{h^{\times}}_{k-k'}(\Omega){}_{k'}\tilde{S}^{\times}(\Omega)\text{,}
\end{equation}
where the superscript $+$ and $\times$ shows which pair of detectors, and we have weighted according to the number of hits of each pair compared to the total contribution.

We may also write the orientation function for the total contribution to the timestream as the weighted sum of the individual detector pairs as
\begin{equation}
    \tilde{h}_n^{\text{tot}} = \frac{N^{+}_{\text{hits}}}{N^{\text{tot}}_{\text{hits}}}\tilde{h}_n^{+} + \frac{N^{\times}_{\text{hits}}}{N^{\text{tot}}_{\text{hits}}}\tilde{h}_n^{\times}.
\end{equation}
Since we are considering a pair differenced signal the temperature signal should have been nominally removed meaning an experiment can attempt to solve for the polarization directly as
\begin{equation}
\begin{pmatrix}\hat{Q}-i\hat{U}\\\hat{Q}+i\hat{U}\end{pmatrix}
=
\begin{pmatrix}\frac{1}{4}\tilde{h}^{\text{tot}}_4&\frac{1}{4}\\\frac{1}{4}&\frac{1}{4}\tilde{h}^{\text{tot}}_{-4}\end{pmatrix}^{-1}
\begin{pmatrix}{}_2\tilde{S}^{d,\text{tot}} \\{}_{-2}\tilde{S}^{d,\text{tot}}\end{pmatrix}\text{.}
\label{eq:Polarisation Calculation Map-make}
\end{equation}
However due to the ``$+\times$'' detector layout there are always the requisite pairs of angles such that $\tilde{h}_{4}=0$.\footnote{We note that this was done implicitly in the map-making approach taken in \cite{Wallisetal2016} and \cite{2020arXiv200800011M} which made assumptions based on the use of a ``$+\times$'' focal plane.} As such we may simply write
\begin{equation}
\begin{pmatrix}\hat{Q}-i\hat{U}\\\hat{Q}+i\hat{U}\end{pmatrix}
=
\begin{pmatrix}0&2\\2&0\end{pmatrix}
\begin{pmatrix}{}_2\tilde{S}^{d,\text{tot}} \\{}_{-2}\tilde{S}^{d,\text{tot}}\end{pmatrix}\text{.}
\label{eq:Polarisation Calculation +x}
\end{equation}

We shall also apply a further simplification for our subsequent calculations, as is done in \cite{Wallisetal2016,2020arXiv200800011M}, that the detector pairs are effectively colocated such that we may set
\begin{equation}
    \tilde{h}_n^{\times}(\Omega) = e^{-in\frac{\pi}{4}} \tilde{h}_n^{+}(\Omega)\text{,}
\end{equation}
where the $e^{-in\frac{\pi}{4}}$ accounts for the rotation of $45^{\circ}$ between the $+$ and $\times$ orientation of the detectors. Provided the detector pairs are located close to one another on the focal plane this should give a fair representation. See section \ref{section:Full Focal Plane} for further information on how to treat a full focal plane.

\subsection{Full TOD Simulation}
\label{section:Full TOD Simulation}
To provide a baseline to compare the map-based simulation method to we run realistic full TOD simulations. These involve looping over the telescope pointing information from the full scanning strategy, incurring a large time cost. We present results for scanning strategies representative of a satellite survey and both ``Deep'' and ``Shallow'' ground-based surveys.

For the satellite case we use the {\it EPIC} scanning strategy, defined according to the parameters in table \ref{tab:scans}, which was designed
to optimise crossing angle coverage. Further details are available in \cite{2008arXiv0805.4207B}.

The ground-based surveys, also defined in table \ref{tab:scans}, are indicative of scanning strategies for an Atacama based instrument performing constant elevation scans at two elevations for 30 days each -- swiping back and forth in azimuth at a constant scan rate. Appendix \ref{section:GBSS} gives more detail. The ``Deep'' surveys are traditionally performed over a smaller patch of sky allowing for more integration time in order to heavily suppress noise for primordial $B$-mode searches. The ``Shallow'' surveys are performed over a wide patch of sky to aid in the characterisation of the lensing signal and other science goals e.g. measuring the Sunyaev–Zeldovich effect from galaxy clusters, or measuring Neutrino masses through high-$\ell$ effects \citep[e.g.][]{2018SPIE10708E..41S,2019JCAP...02..056A}.

\begin{table}
	\centering
	\caption{Details of the scanning strategy parameters used in the simulations for both the {\it EPIC} satellite and the ``Shallow'' and ``Deep'' ground-based surveys.}
	\label{tab:scans}
	\begin{tabular}{lcc} 
		\hline
		{\it EPIC} Satellite Scan\\
		\hline
		NSIDE & 2048\\
		Beam Size & 7 arcmin\\
		Boresight Angle & $50^{\circ}$\\
		Precession Angle & $45^{\circ}$\\
		Spin Period & 1 min\\
		Precession Period & 3 hours\\
		$f_{samp}$ & 500 Hz\\
		Survey Time & 1 year\\
		\hline
		Ground-Based Scans\\
		\hline
		NSIDE & 256\\
		Beam Size & 30 arcmin\\
		$f_{samp}$ & 10 Hz\\
		Scan Rate & $1^{\circ}$/s\\
		Observatory Location & Atacama\\
		Survey Time & 30 days\\
		Elevations & $35^{\circ}$ and $55^{\circ}$\\
		``Deep'' Azimuth Range Rising (Setting) - &  \\Elevation=$35^{\circ}$ & $111^{\circ}$ to $136^{\circ}$ ($-111^{\circ}$ to $-136^{\circ}$) \\ Elevation=$55^{\circ}$ & $110^{\circ}$ to $148^{\circ}$ ($-110^{\circ}$ to $-148^{\circ}$) \\
		``Shallow'' Azimuth Range Rising (Setting) - & \\ Elevation=$35^{\circ}$ & $53^{\circ}$ to $119^{\circ}$ ($-53^{\circ}$ to $-119^{\circ}$)\\ Elevation=$55^{\circ}$ & $72^{\circ}$ to $117^{\circ}$ ($-72^{\circ}$ to $-117^{\circ}$)
	\end{tabular}
\end{table}

As mentioned, we use a simple focal plane as in \cite{Wallisetal2016}, where two detector pairs are considered oriented at $45^{\circ}$ with respect to each other. We choose not to include noise to ensure the comparison of the systematic effects generated from the full TOD approach to those of the map-based simulation methods is clear. Since correlation between random noise and the systematic effects considered here is not expected, our conclusions regarding the ability of the map-based simulation to mimic the full TOD approach are robust to this choice. White noise can be trivially included into the map-based simulations by simply adding noise maps to the measured data as detailed in e.g. \cite{1997PhRvD..56.4514T,1997ApJ...480L..87T}.

As an input to the TOD simulation CMB maps of $I$, $Q$, and $U$ are created using the {\it SYNFAST} routine of the {\it HEALPIX} package \citep{2005ApJ...622..759G}. A 6-parameter $\Lambda$CDM cosmology was used to generate the input CMB power spectra, using the best fitting 2015 {\it Planck} results specified by the following cosmological parameter values: Hubble constant -- $H_{0} = 67.3$, Baryon density -- $\Omega_{b} = 0.0480$, Dark matter density -- $\Omega_{cdm} = 0.261$, Optical depth to reionization -- $\tau = 0.066$, Scalar spectral index -- $n_{s} = 0.968$, Amplitude of scalar perturbations -- $A_{s} = 2.19\times10^{-9}$  \citep{2016A&A...594A..13P}. The Boltzmann code CLASS \citep{2011JCAP...07..034B} was used to generate the input spectra.

The input maps include both primordial $B$-modes at a level of $r = 0.001$ and $B$-modes induced by gravitational lensing (approximated as Gaussian). The parameters used to generate the scan and input maps are listed in table \ref{tab:scans}. In the simulations of differential gain and pointing the input maps were convolved with a Gaussian beam. In the case of differential ellipticity a different convolution process was implemented -- see section \ref{section:EllipTOD} for further details.

The TOD samples are generated from a set of pointings given by a scanning strategy. In the absence of systematics the four detectors used each measure a signal according to the detector equation of
\begin{equation}
    d^{X}_{i} = I(\Omega) + \frac{1}{2}((Q-iU)(\Omega)e^{2i\psi_i^X} + (Q+iU)(\Omega)e^{-2i\psi_i^X}) \text{,}
    \label{eq:Detector Equation}
\end{equation}
where $i \in 1,2$ denotes the pair considered, $X \in A,B$ denotes the detector within a pair, and $\psi_i^X$ is the combined crossing angle of the instrument $\psi$ and angle offset of the individual detector from this $\alpha_i^X$ as $\psi_i^X = \psi + \alpha_i^X$. The $\alpha_i^X$ values are offset by $90^{\circ}$ between the two detectors $A$ and $B$ in a pair, and there is $45^{\circ}$ offset between the pairs 1 and 2. For simplicity we assume that $\alpha_i^A = 0$ such that the reference frame of detector $d^A_1$ is aligned with the instrument i.e. $\psi_1^A = \psi$.

We difference the detectors within a pair to produce differenced timestreams. The timestreams are then subjected to simple binned map-making according to equation \ref{eq:Polarisation Calculation Map-make} to solve for the polarization signals -- Stokes $\hat{Q}$ and $\hat{U}$.

\subsubsection{Gain Systematic}
Systematics are injected at the detector time stream level. A differential gain systematic is included by adding some $g_{i}$ offset due to gain miscalibration to each detector as
\begin{equation}
    d_{i}^{X} = (1+g_{i}^{X}) (I(\Omega) + \frac{1}{2}((Q-iU)(\Omega)e^{2i\psi_{i}^{X}} + (Q+iU)(\Omega)e^{-2i\psi_{i}^{X}})) \text{.}
    \label{eq:Gain Single Detector}
\end{equation}
Since we are considering a pair-differencing experiment, the differenced signal for a single pair is then given by
\begin{equation}
\begin{split}
    S_i &= \frac{1}{2}[d^A_i - d^B_i]
    \\
    &=\frac{1}{2}\bigg[(g^A_i-g^B_i)I(\Omega)
    + \frac{1}{2}\Big((Q-iU)(\Omega)[2+g^A_i+g^B_i]e^{2i\psi_i^A}
    \\
    &+ (Q+iU)(\Omega)[2+g^A_i+g^B_i]e^{-2i\psi_i^A}\Big)\bigg]\text{.}
\end{split}
\end{equation}
The gain simulations we present use systematics levels of $1\%$ differential gain ($\delta g_i = g^A_i-g^B_i$ = 0.01). We apply this by setting $g^A_i=0.01$ and $g^B_i=0$ for each pair of detectors included in the simulation.

\subsubsection{Pointing Offset}
A pointing offset is included by adding some misalignment of magnitude $\rho^X_{i}$ in direction $\chi^X_{i}$ to each detector as
\begin{equation}
\begin{split}
d_i^X &= I(\Omega) + \, \frac{1}{2}(e^{2 i \, \psi_i^X} \, (Q-iU)
    + \, e^{-2 i \, \psi_i^X} \, (Q+iU))
    \\
    &+ \frac{\rho^X_i}{2} \, e^{i \, (\psi_i^A + \chi^X_i)} \, \bar{\eth} I
    + \frac{\rho^X_i}{2} \, e^{-i \, (\psi_i^A + \chi^X_i)} \, \eth I\text{,}
\end{split}
\end{equation}
where we only include the temperature leakage component in this case. Here we have used the spin raising operator $\eth = \frac{\partial }{\partial y}+i\frac{\partial }{\partial x}$ and its conjugate (the spin lowering operator, denoted by a bar).

For consistency with the previous conventions adopted in \cite{Wallisetal2016,2020arXiv200800011M} we have adopted the reference frame of the orientation angle $\psi_i^A$ of the $d^A_i$ detector in each pair to define the $\rho^X_{i}$ and $\chi^X_{i}$ parameters. This means that the detectors within each pair share a common reference frame when describing the pointing offset meaning there is no factor of $\pi/2$ in the pointing offset of the second detector. However this reference frame is rotated through $\pi/4$ between the two pairs we consider $+ \rightarrow \times$. We reiterate that this is just a choice of convention, and this specific choice is not required to apply the methods we present.

In the pair differenced case this results in a contribution by a single pair of
\begin{equation}
\begin{split}
S_i = \frac{1}{2}[d^A_i - d^B_i]
    &=\frac{1}{2} \, e^{2 i \, \psi^A_i} \, (Q-iU)
    + \frac{1}{2} \, e^{-2 i \, \psi^A_i} \, (Q+iU)
    \\
    &+ \frac{\zeta_{i}}{4} \, e^{i \, \psi^A_i} \, \bar{\eth} I
    + \frac{\zeta^*_{i}}{4} \, e^{-i \, \psi^A_i} \, \eth I\text{,}
\end{split}
\end{equation}
where $\zeta = \rho^{A}e^{i\chi^A} - \rho^{B}e^{i\chi^B}$ gives the effect of the pointing mismatch. In this case we only consider the $I\rightarrow P$ leakage induced by the differential pointing error and ignore the smaller $P\rightarrow P$ terms \citep{2020arXiv200800011M}. The {\it HEALPY} routine {\it alm2map\_der1} \citep{2005ApJ...622..759G,Zonca2019} was used to generate the first derivatives of the temperature fields to include the differential pointing effects. This is a similar approach to how lensing is incorporated in simulations in \cite{2013JCAP...09..001N}.\footnote{An alternative would be to directly incorporate the offset during the TOD generation as in e.g. \cite{2020MNRAS.491.1960T}. This may alter the corresponding results slightly due to complications arising from Healpix interpolation effects that would need to be accounted for. By dealing directly with the Taylor expanded quantities ($\eth I$ etc.) we avoid this complication.}

The pointing simulations use a systematic level of $\rho^A_{1} = \rho^A_{2} = 0.1$ arcmin, $\rho^B_{1} = \rho^B_{2} = 0.0$ arcmin, and $\chi_i^X=0$ radians for all detectors. These levels of systematic are indicative of differential systematics seen in recent CMB ground-based surveys \cite[e.g.][]{2015ApJ...814..110B,1403.2369}.

\subsection{Application of the map-based technique to a Gain Systematic}
\begin{figure*}
  \centering
  \includegraphics[width=2\columnwidth]{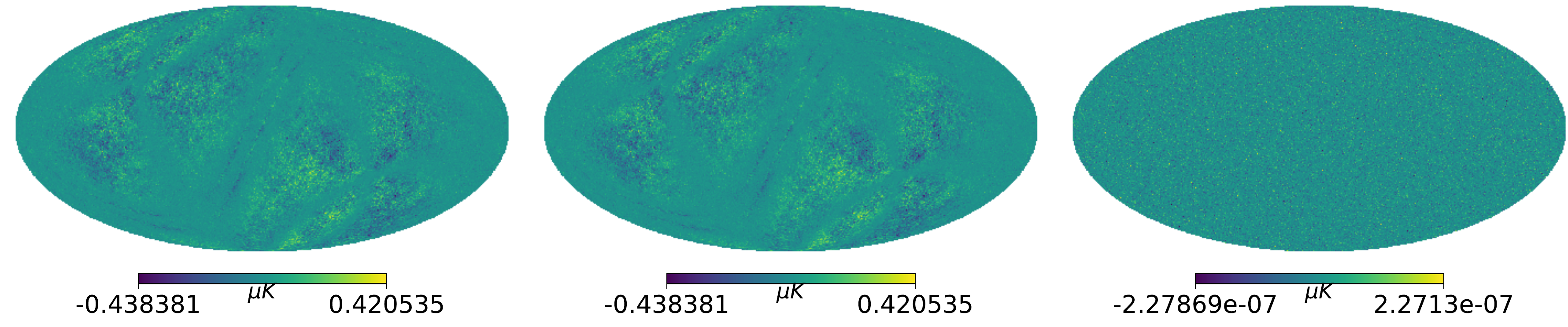}
  \includegraphics[width=2\columnwidth]{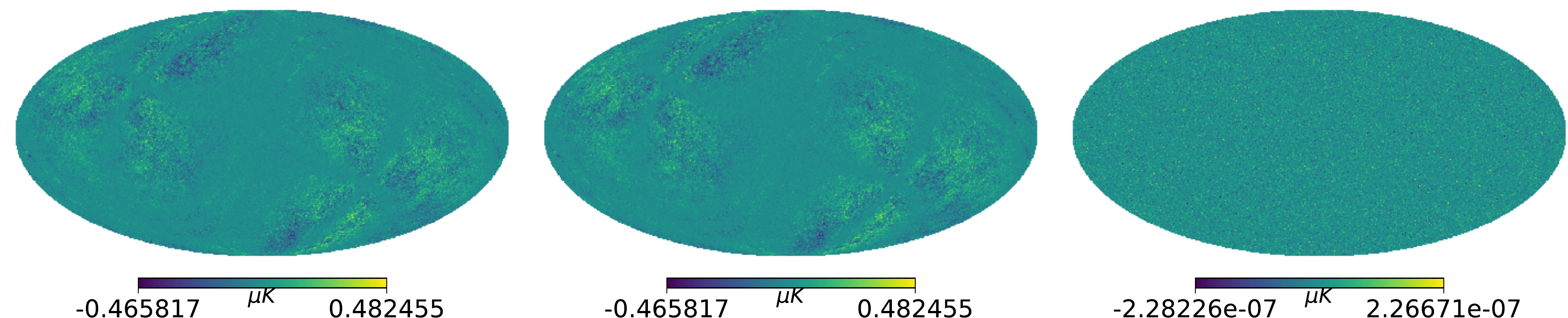}
\caption{Gain systematic field -- created by taking the difference between the output map containing systematics and the input map. Top row - Q. Bottom row - U. Left column - TOD simulation using {\it EPIC} scan. Middle column - Map-based simulation with $\tilde{h}_n$ maps created using {\it EPIC} scan. Right column - Residual between the first two columns. The low level of residual shows there is a good agreement between the full TOD and map-based method. The track like structure seen in the first two columns is a result of structure in the EPIC scanning strategy resulting in better crossing angle coverage and a reduced systematic in certain areas. The map-based approach captures these features of the full TOD simulation well.}
\label{figure:hnmapgain}
\end{figure*}
\begin{figure}
  \centering
  \includegraphics[width=\columnwidth]{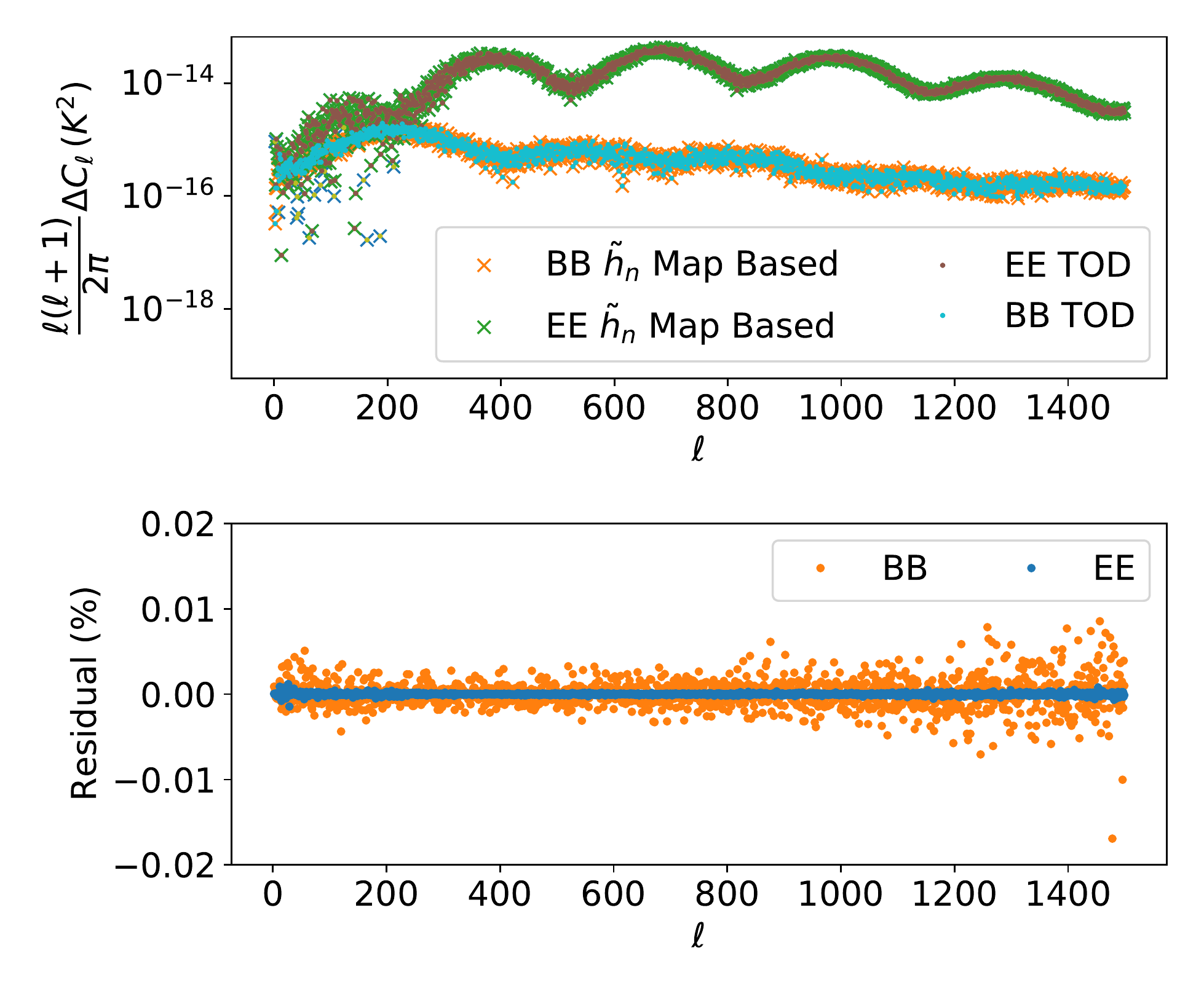}
\caption{Power spectrum of the gain systematic -- $\Delta C_{\ell}$ indicates it is the difference between a simulation containing the systematic and a simulation with no systematic present leaving just the systematic leaked signal. The full {\it EPIC} TOD and map-based simulation match well including the scatter for both the $B$-mode and $E$-mode -- note the low levels of residual between the two approaches in the lower panel.}
\label{figure:EpicGainCls}
\end{figure}
\begin{figure*}
  \centering
  \includegraphics[width=2\columnwidth]{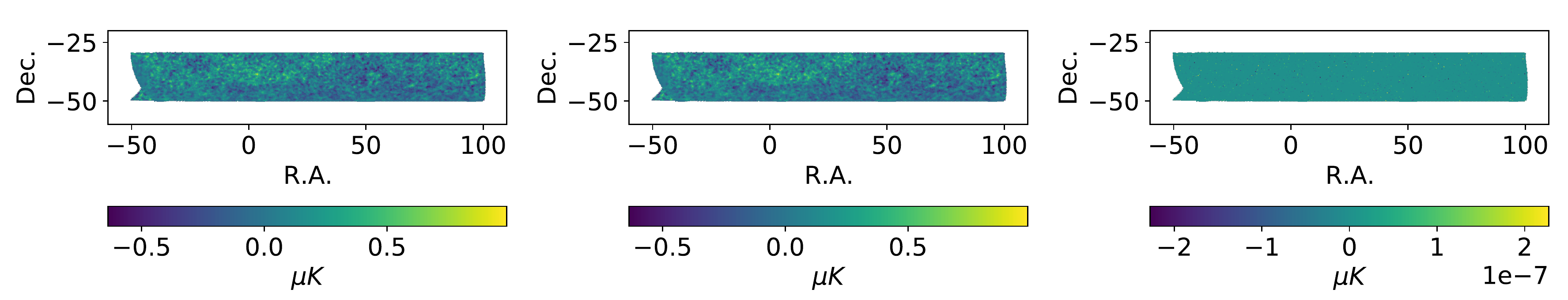}
  \includegraphics[width=2\columnwidth]{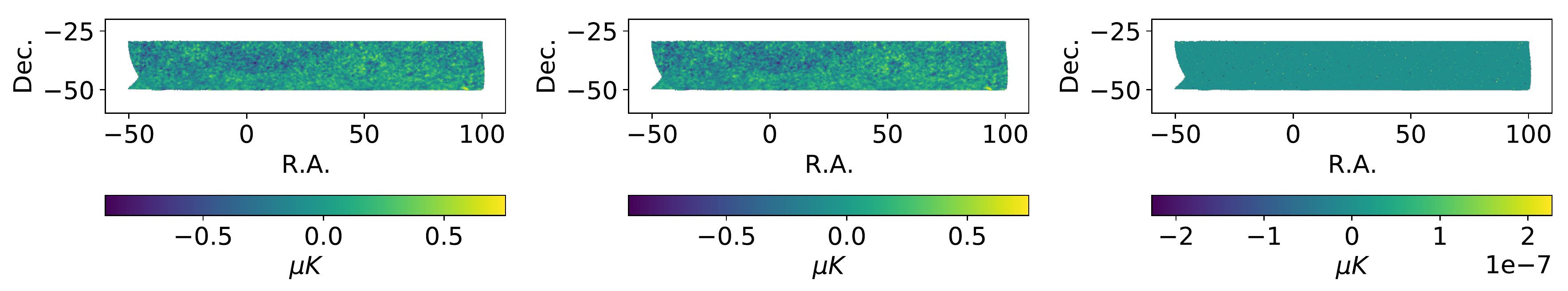}
\caption{Gain systematic field -- created by taking the difference between the output of a systematic containing simulation and the input map. Top row - Q. Bottom row - U. Left column - TOD simulation using the ``Deep'' ground-based scan. Middle column - Map-based simulation with $\tilde{h}_n$ maps created using the ``Deep'' scan. Right column - Residual between the first two columns which shows there is a good agreement between the two methods. The map-based approach captures the features of the full TOD simulation well.}
\label{figure:DeepGainMaps}
\end{figure*}
\begin{figure}
  \centering
  \includegraphics[width=\columnwidth]{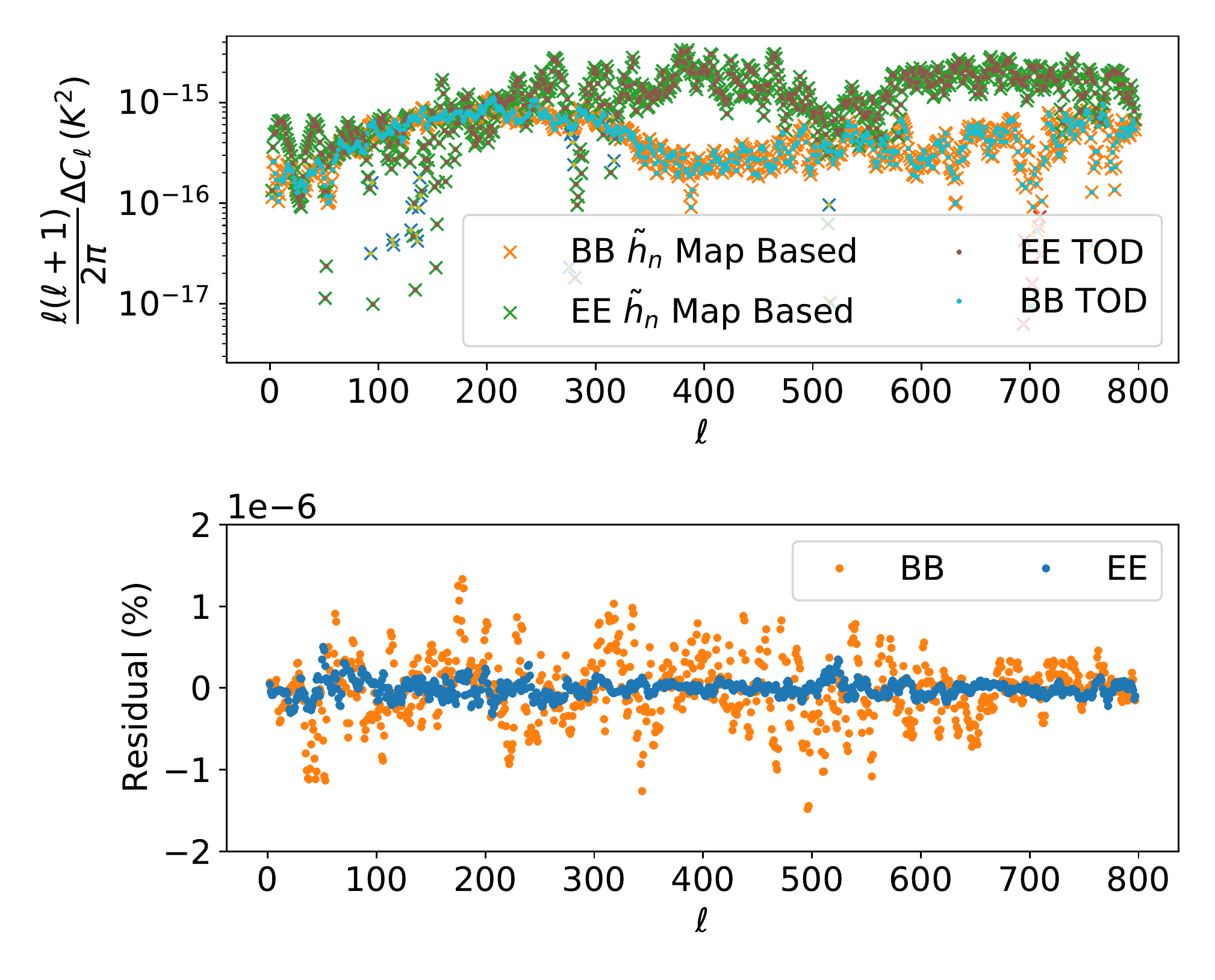}
\caption{Pseudo power spectrum of the gain systematic -- $\Delta C_{\ell}$ indicates it is the difference between a simulation containing the systematic and a simulation with no systematic present leaving just the systematic leaked signal. The ``Deep'' ground-based TOD and map-based simulation match well including the scatter present as is evidenced by the low level of the residuals in the lower panel.}
\label{figure:DeepGainCls}
\end{figure}

The effect of differential gain arises due to a gain mismatch between detectors in differenced pairs. This leads to leakage of the much larger temperature signal into the polarization, and a direct amplification of the polarization signal itself also remains. To match the full TOD setup we are comparing to, we consider two pairs of differenced detectors oriented at $45^{\circ}$ to one another. Following the process outlined in section \ref{section:Map-Based Sims} and applying equations \ref{eq:spinsyst} and \ref{eq:Polarisation Calculation +x}, this leads to a total polarization signal of
\begin{equation}
\begin{split}
&(\hat{Q}+i\hat{U})(\Omega) = \tilde{h}_0(\Omega) (Q+iU)(\Omega)
    + \frac{1}{2} \tilde{h}_2(\Omega) (\delta g_1 - i\delta g_2) \, I(\Omega)
    \\
    &+ \frac{1}{4}\tilde{h}_0(\Omega)(g^A_1 + g^B_1 + g^A_2 + g^B_2) \, (Q+iU)(\Omega)
    \\
    &+ \frac{1}{4}\tilde{h}_4(\Omega)(g^A_1 + g^B_1 - g^A_2 - g^B_2) \, (Q-iU)(\Omega) \text{,}
    \label{eq:Map Based Differential Gain}
\end{split}
\end{equation}
where $g^X_i$ is the gain offset of detector $X \in A,B$ of pair $i \in 1,2$ due to a gain miscalibration, and $\delta g_i = g^A_i-g^B_i$ is the gain mismatch between detector $A$ and the detector perpendicular to it $(B)$ in the pair. Equation \ref{eq:Map Based Differential Gain} provides the necessary information to implement the map-based simulation of differential gain.

\begin{figure*}
  \centering
  \includegraphics[width=2\columnwidth]{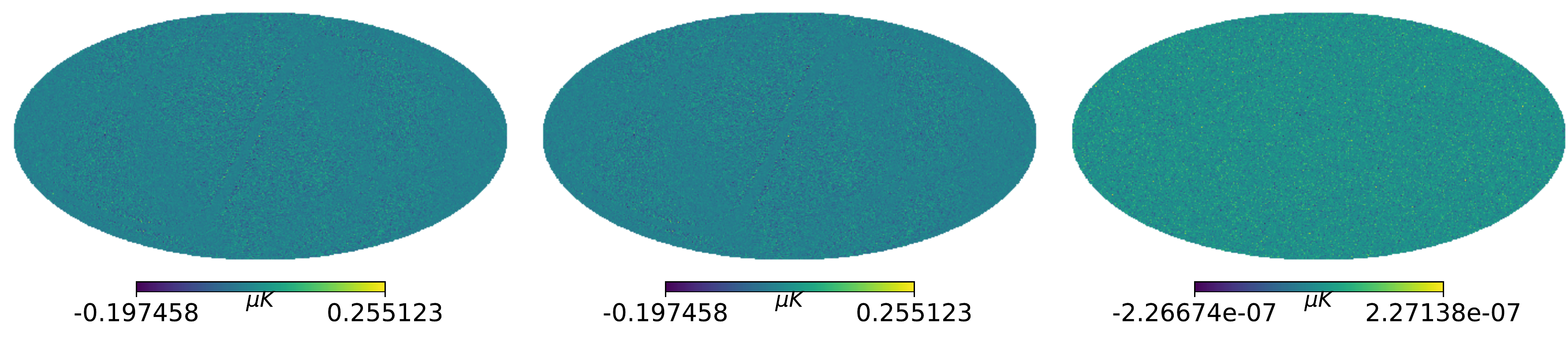}
  \includegraphics[width=2\columnwidth]{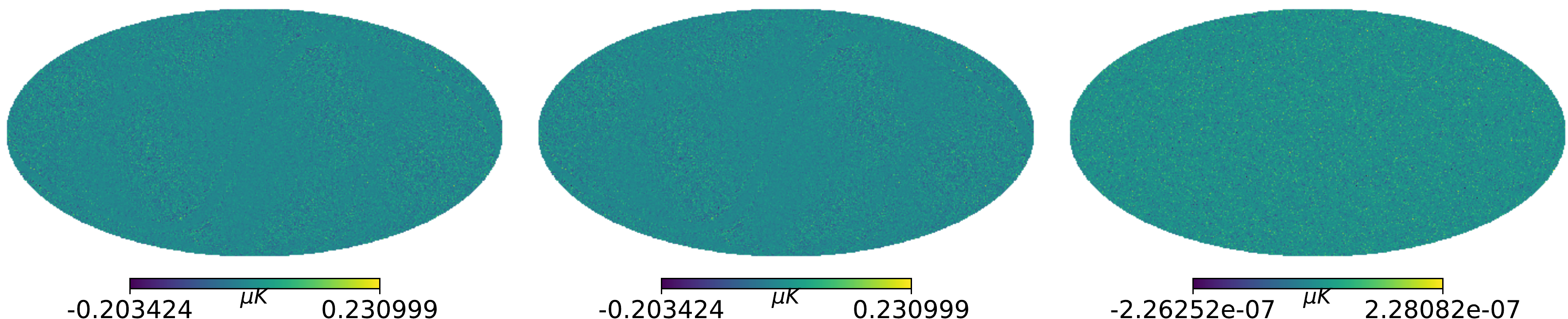}
\caption{Pointing systematic field -- created by taking the difference between the systematic containing output map and the input map. Top row - Q. Bottom row - U. Left column - TOD simulation using the {\it EPIC} scan. Middle column - Map-based simulation with $\tilde{h}_n$ maps created using the {\it EPIC} scan. Right column - Residual between the first two columns which is low showing there is good agreement between the two approaches. The track like structure seen in the first two columns is a result of structure in the {\it EPIC} scanning strategy resulting in better crossing angle coverage in those areas and thus a reduced systematic. The map-based approach captures these features of the full TOD simulation well.}
\label{figure:hnmappointing}
\end{figure*}
\begin{figure}
  \centering
  \includegraphics[width=\columnwidth]{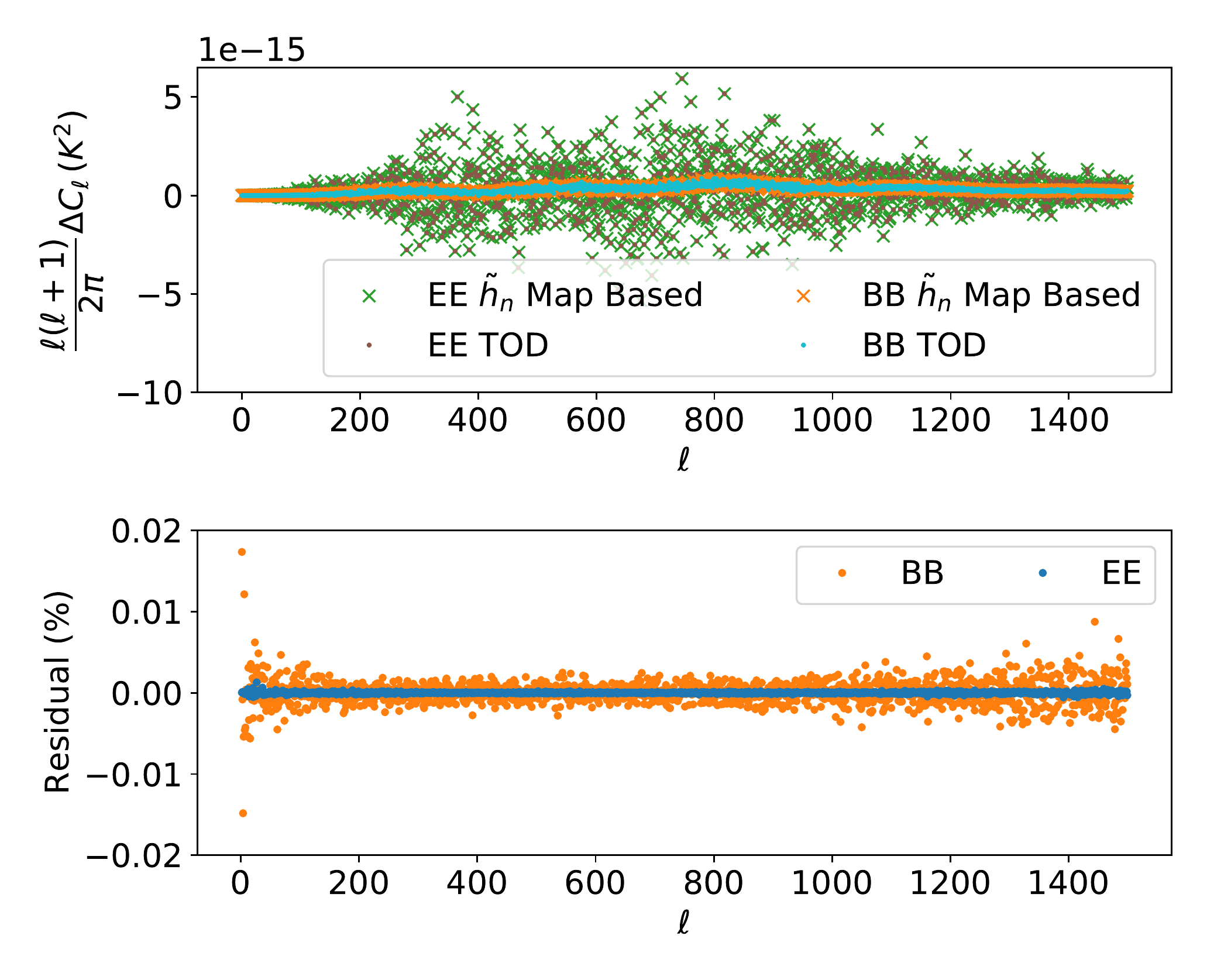}
\caption{Power spectrum of the pointing systematic (temperature leakage only) -- $\Delta C_{\ell}$ indicates it is the difference between a simulation containing the systematic and a simulation with no systematic present leaving just the systematic leaked signal. The full {\it EPIC} TOD and map-based approach match well including the scatter. The low levels of the residuals demonstrate good agreement between the two methods.}
\label{figure:EpicPointingCls}
\end{figure}

We provide comparisons of the map-based simulation approach to full TOD simulations for both a satellite survey and ground-based survey. In figures \ref{figure:hnmapgain} and \ref{figure:DeepGainMaps} we show the results for a gain systematic for both a full TOD simulation and the map-based approach for the {\it EPIC} satellite scan and the ``Deep'' ground-based scan respectively. We plot the contribution of the systematic to the polarization maps by taking the difference between the output of the simulations including a gain systematic and the input map. We see very close agreement between the TOD simulation and the map-based approach for both the satellite scan and the ground based scan -- this is demonstrated by the low level of residual between the two approaches. Particularly evident in the satellite plots is the track like structure which is a result of structure in the {\it EPIC} scanning strategy. Satellite missions have access to a larger range of crossing angles than ground-based experiments which results in the evident reduction in the systematic. The map-based approach captures these features of the full TOD simulation well.

In figures \ref{figure:EpicGainCls} and \ref{figure:DeepGainCls} we also show the resulting power spectra contribution of the gain systematic in the case of the full TOD and map-based simulation. In the satellite case (figure \ref{figure:EpicGainCls}) no mask is applied and we plot the full sky power spectra, whereas in the ground-based case (figure \ref{figure:DeepGainCls}) it is the pseudo-spectra plotted, with the effects of the mask still present. In this case we plot $\Delta C_{\ell}$ which is the difference between the simulation containing the systematic and a simulation with no systematic present -- this isolates the systematic contribution from the true signal. We see good agreement between the full TOD and fast map-based simulation with the map-based simulation even capturing the scatter on the data which stems in part from cosmic variance and in part from the scanning strategy. The ability to retain this scatter sets this map-based method apart from analytic models of the systematic power spectra \cite[e.g.][]{2020arXiv200800011M} as it retains information specific to both the CMB realisations and the scanning strategy -- this means the method is appropriate for use with monte-carlo simulations of CMB surveys including systematic effects. We further elucidate this agreement by plotting the residual between the two approaches which is shown to be small at all multipoles considered.

\begin{figure*}
  \centering
  \includegraphics[width=2\columnwidth]{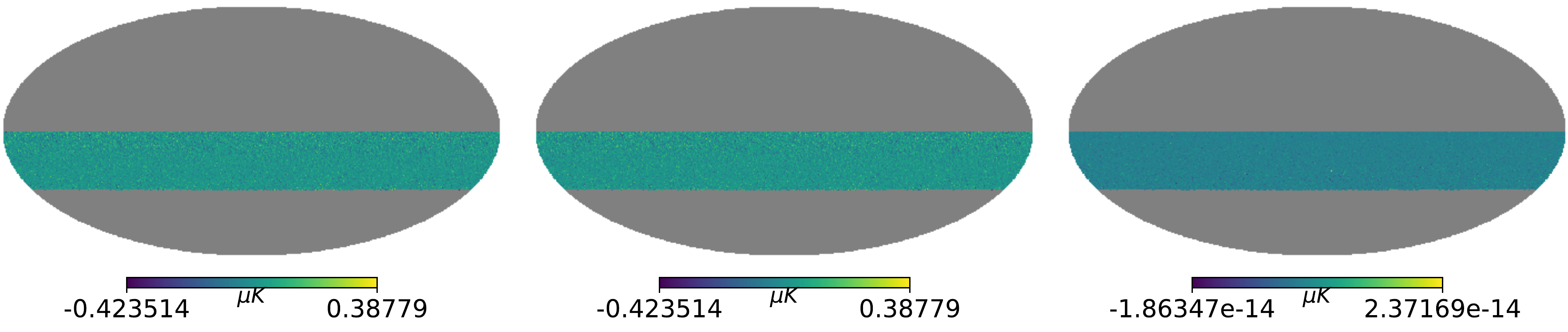}
  \includegraphics[width=2\columnwidth]{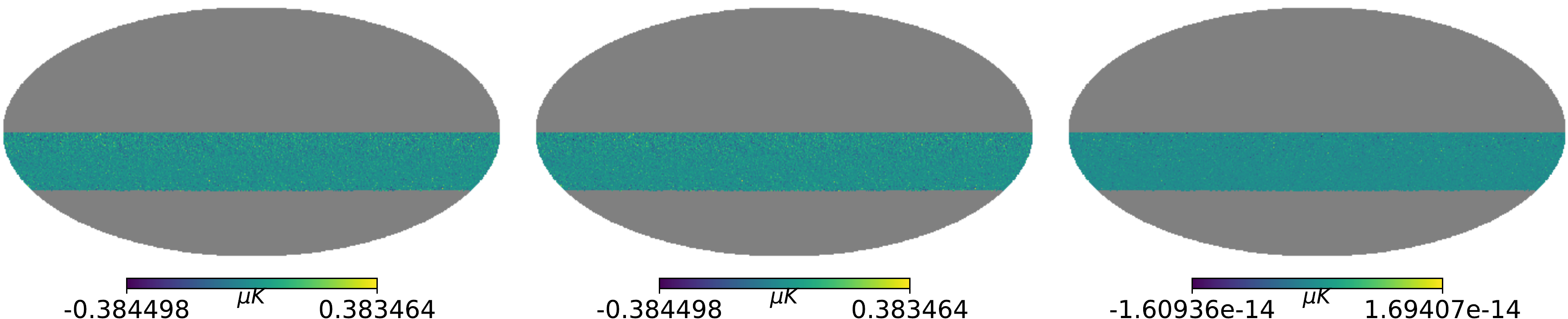}
\caption{Pointing systematic field -- created by taking the difference between the systematic containing output map and the input map. Top row - Q. Bottom row - U. Left column - TOD simulation using the ``Shallow'' ground-based scan. Middle column - Map-based simulation with $\tilde{h}_n$ maps created using the ``Shallow'' scan. Right column - Residual between the first two columns which shows there is a good agreement between the two approaches. The map-based method captures the features of the full TOD simulation well.}
\label{figure:ShallowPointingMaps}
\end{figure*}
\begin{figure}
  \centering
  \includegraphics[width=\columnwidth]{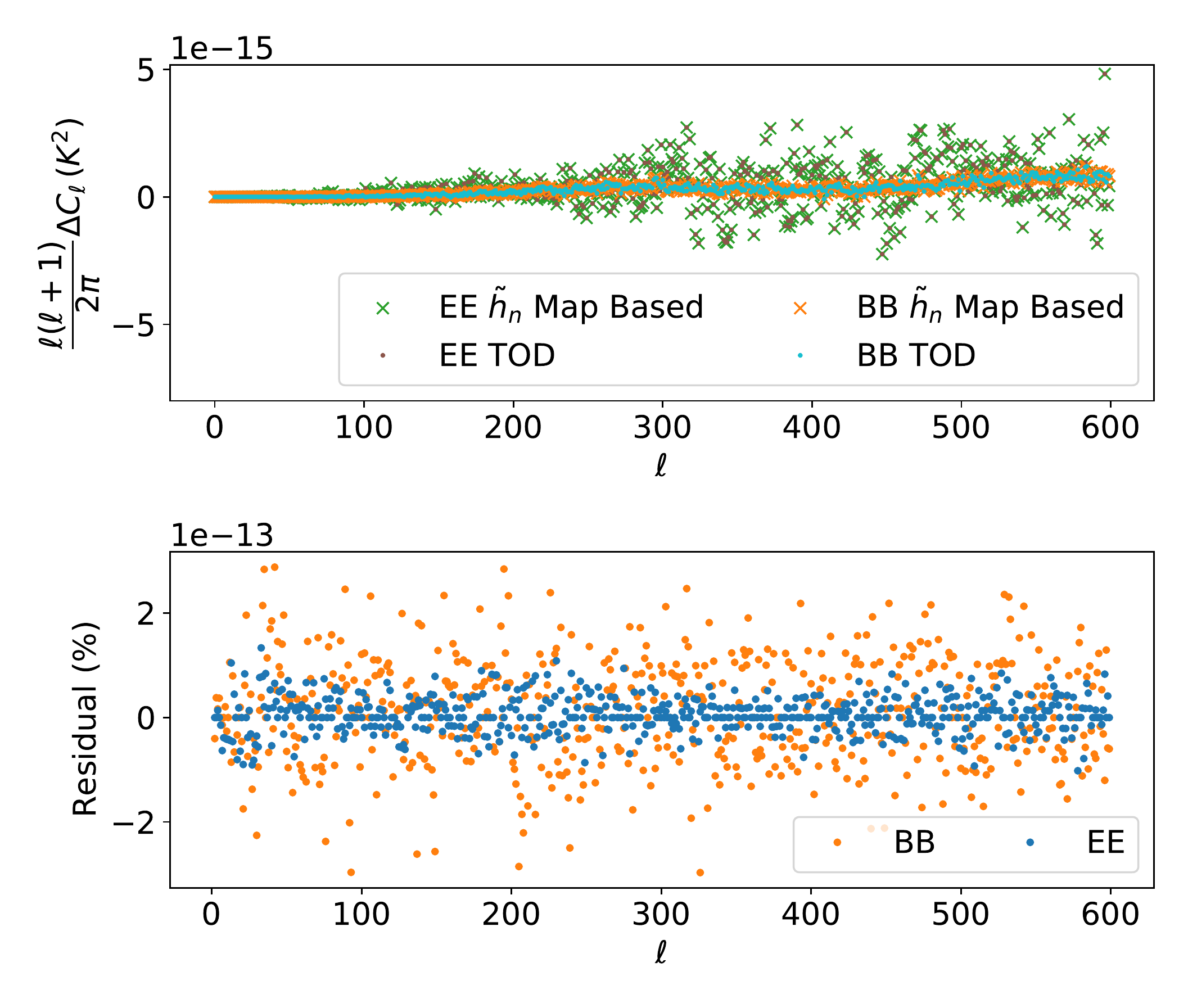}
\caption{Pseudo power spectrum of the pointing systematic -- $\Delta C_{\ell}$ indicates it is the difference between a simulation containing the systematic and a simulation with no systematic present leaving just the systematic leaked signal. The ``Shallow'' ground-based TOD and map-based simulation match well including the scatter present. The small residuals demonstrate the precise agreement between the two approaches. In the case of the $EE$ spectrum the discrepancy between the two sets of points is close to machine precision in many cases (i.e. the residuals are exactly zero). The $EE$ residuals are thus dominated by artificial effects arising from the finite machine precision -- hence the ``quantized''-like appearance in the blue points.}
\label{figure:ShallowPointingCls}
\end{figure}
\subsection{Application of the map-based technique to a Pointing Systematic}
The misalignment of the beams of two detectors within a pair will result in a differential pointing systematic. In order to match the full TOD setup we are comparing to, we consider two pairs of differenced detectors oriented at $45^{\circ}$ to one another, hence the factors of $e^{-in\pi/4}$ for the second pair in the pointing offset terms. Again following the process outlined in section \ref{section:Map-Based Sims} and applying equations \ref{eq:spinsyst} and \ref{eq:Polarisation Calculation +x} leads to a total polarization signal of
\begin{equation}
\begin{split}
&(\hat{Q}+i\hat{U})(\Omega)
    \\
    &= \tilde{h}_0 (Q+iU)(\Omega)
    + \frac{1}{4}\tilde{h}_1(\rho^A_1 \, e^{i \, \chi^A_1} + \rho^A_2 \, e^{i \, (\chi^A_2 - \pi/4)}) \, \bar{\eth} I(\Omega)
    \\
    &+ \frac{1}{4}\tilde{h}_3(\rho^A_1 \, e^{-i \, \chi^A_1}
    + \rho^A_2 \, e^{-i \, (\chi^A_2 + 3\pi/4)}) \eth I(\Omega)
    \label{eq:Differential Pointing}
\end{split}
\end{equation}
where in this case we are only considering the $I\rightarrow P$ leakage induced by differential pointing error and ignore the smaller $P\rightarrow P$ terms \citep{2020arXiv200800011M}. In this case by combining maps of $\tilde{h}_1$, $\tilde{h}_3$, $\eth I(\Omega)$ and $\bar{\eth} I(\Omega)$, we arrive at a map that represents the spurious signal due to the systematic, without requiring a full TOD simulation.\footnote{Note that equation \ref{eq:Differential Pointing} corrects a typo in \cite{Wallisetal2016} and \cite{2020arXiv200800011M}.}

As for the previous systematic we provide comparisons between the map-based simulation approach and full TOD simulations for both a satellite survey and ground-based survey. In figures \ref{figure:hnmappointing} and \ref{figure:ShallowPointingMaps} we show the results for a differential pointing systematic for both a full TOD simulation and the map-based approach for the {\it EPIC} satellite scan and the ``Shallow'' ground-based scan respectively. We again plot the contribution of the systematic to the polarization maps by taking the difference between the output of the simulations including a gain systematic and the input map. We see close agreement between the TOD simulation of the differential pointing systematic and its map-based counterpart. As in the case of gain previously, we see that there is a track like structure in the satellite case which is a result of structure in the {\it EPIC} scanning strategy. The structure emanates from the scanning strategy and the inclusion of the $\tilde{h}_1$ and $\tilde{h}_3$ quantities in the map-based approach mean it is captured well.

Figures \ref{figure:EpicPointingCls} and \ref{figure:ShallowPointingCls} show the resulting power spectra contribution of the differential pointing systematic in the case of the full TOD and map-based simulation. Figure \ref{figure:EpicPointingCls} shows the satellite case where full sky power spectra are plotted with no mask applied, whereas figure \ref{figure:ShallowPointingCls} shows the ground-based case where the pseudo-spectra are plotted, with the effects of the mask still present. We again plot $\Delta C_{\ell}$ which is the difference between the simulation containing the systematic and a simulation with no systematic present which isolates the systematic contribution from the true signal. The full TOD and map-based simulation results clearly agree well. The map-based simulation also captures the scatter on the data which stems in part from cosmic variance and in part from the scanning strategy. The residual between the two approaches is small at all multipoles considered showing the map-based approach works well.

\subsection{Summary and explanation of timings}
We have demonstrated that the map-based simulations provide a viable approach to simulate systematics through comparison to full TOD simulations. By using stored average scan quantities we can achieve a massive speed up in comparison to the full TOD simulation whilst retaining effects of the scanning strategy.

The speed of a full TOD simulation will depend on the scan strategy. In particular, the sampling frequency, and the length of observation will dictate how much data the simulation will be required to loop over. However, this is not the limiting factor for the map-based approach which makes use of the stored $\tilde{h}_n$ scan strategy maps to remove the time dimension of the simulation, meaning only a single calculation is needed for each pixel in map-space.

Performing the {\it EPIC} TOD simulation for a 1 year scan with the parameters shown in table \ref{tab:scans} takes 10--15 hours of CPU time. The map-based simulation by comparison takes $\sim$10 seconds, i.e. the map-based simulation is $\sim$3600--5400 times faster.\footnote{Given that the scan parameters are not the limiting factor for the map-based approach then the speed up for longer TOD sets would be even greater.} The significant speed up offered by this approach could be used in tandem with monte-carlo techniques to quickly simulate systematics in CMB surveys including the effects of the scanning strategy. This will speed up detailed forecasts of the effects of a number of systematics on upcoming surveys. This speed up also extends to realistic numbers of detectors, as will be shown in section \ref{section:Full Focal Plane}, which is an important capability given the many thousands of detectors expected in future focal plane setups.

\section{Constant Elevation Scan Approach}
\label{section:CES Approach}
In order to perform the map-based simulations in the previous sections we utilised the simplification of equation \ref{eq:RHS Simplification}. This was possible as we were dealing with signals which did not vary with the orientation angle i.e. $I$, $Q$, and $U$, signals smeared by a circular beam. However in some cases these signals can vary with crossing angle -- e.g. when smeared by an elliptical beam. In this case equation \ref{eq:RHS Simplification} no longer holds -- the observed signal which we wrote earlier as $d_j = \sum_{n\geq0}(d_n^Q\cos(n\psi_j)+d_n^U\sin(n\psi_j))$ no longer has well defined spin dependence due to the signals $d_n^Q(\psi_j)$ and $d_n^U(\psi_j)$ now depending on the crossing angle $\psi_j$ -- and we must simulate these quantities using a different approach. In the case of ground-based strategies we can utilise restrictions that are imposed on the scanning strategy to facilitate an alternative fast method, that imitates the full TOD well whilst also dealing with the dependence of the signals on the crossing angle.

There are a number of factors that influence scan strategy design, for example control of elevation-dependent and azimuth-dependent contributions to the system such as the atmosphere and ground pickup. These considerations usually dictate that ground-based CMB surveys should use CESs. As discussed in \cite{2021arXiv210202284T}, the use of CESs imposes a fundamental limit on the range of crossing angles that can be achieved in each pixel. Specifically the use of CESs means that each scan at each elevation can contribute only two distinct crossing angles per pixel -- one from when the observation field is rising and one from when it is setting, with the further restriction that $\psi^{\text{set}}=-\psi^{\text{rise}}$.\footnote{The crossing angle will be roughly constant across the extent of a sky pixel provided the pixels are small enough (NSIDE $\gtrsim 128$). See \cite{2021arXiv210202284T} for further discussion on this requirement.} We shall refer to this scanning approach as ``NERS'' as we are using N elevations measured at rising and setting.

We may utilise the constraints of NERS to develop an approach to simulate ground-based systematics quickly with scan information encoded. Since NERS dictates the crossing angle coverage allowed in each pixel we can develop an approach capable of dealing with systematics that are allowed to vary with crossing angle -- i.e. do not have a well defined spin-dependence.

The quantities from the right hand side vector of the map-making equation (equation \ref{eq:3x3 hn map making}) are given by
\begin{equation}
    \langle d_j e^{ik\psi_j} \rangle = \frac{1}{N^{\text{tot}}_{\text{hits}}} \sum_{j} d_j e^{ik\psi_j}
\end{equation}
which in this case, using the constraints imposed by NERS, we may write as a sum weighted by the contributions of data taken while setting and rising at each elevation as
\begin{equation}
\begin{split}
    \langle d_j e^{ik\psi_j} \rangle = \frac{1}{N^{\text{tot}}_{\text{hits}}} \sum_{E}^{N_{\text{Elevations}}} \bigg[ &N^{E,\text{rise}}_{\text{hits}} d(\psi^{E,\text{rise}}) e^{ik\psi^{E,\text{rise}}} 
    \\
    &+ N^{E,\text{set}}_{\text{hits}} d(-\psi^{E,\text{rise}}) e^{-ik\psi^{E,\text{rise}}}\bigg]
    \label{eq:j dependent map-based sim}
\end{split}
\end{equation}
where $N_{\text{Elevations}}$ is the number of elevations included in the survey, and $E$ denotes which elevation you are considering. In this case we shortcut the TOD simulation by using stored maps of the number of hits from the rising $(N^{E,\text{rise}}_{\text{hits}})$ and setting $(N^{E,\text{set}}_{\text{hits}})$ part of the survey at each elevation, maps of the two crossing angles allowed in each pixel per elevation -- rising $(\psi^{E,\text{rise}})$ and setting $(\psi^{E,\text{set}})$, and the signal as a function of crossing angle $d_j=d(\psi_j)=\sum_{n\geq0}(d_n^Q(\psi_j)\cos(n\psi_j)+d_n^U(\psi_j)\sin(n\psi_j))$ evaluated at $\psi^{E,\text{rise}}$ and $\psi^{E,\text{set}}$. $d(\psi_j)$ is the observed signal which will be contributed to by both on-sky signals and systematic leaked signals.

We may then combine this with the map-making equation (equation \ref{eq:3x3 hn map making}) or in the case where we solve for just polarization as
\begin{equation}
\begin{pmatrix}\hat{Q}-i\hat{U}\\\hat{Q}+i\hat{U}\end{pmatrix}
=
\begin{pmatrix}\frac{1}{4}\tilde{h}^{\text{tot}}_4&\frac{1}{4}\\\frac{1}{4}&\frac{1}{4}\tilde{h}^{\text{tot}}_{-4}\end{pmatrix}^{-1}
\begin{pmatrix}\langle d_j e^{2i\psi_j} \rangle \\\langle d_j e^{-2i\psi_j} \rangle\end{pmatrix}
\label{eq:polarization j dependent}
\end{equation}
to calculate the effect on the observed signals. Note that in the demonstration presented in section \ref{section:EllipTODvsMBS} we use a ``$+\times$'' focal plane element setup so in that case the diagonals would be set to zero as in section \ref{section:+XSimplification}.

One can use equations \ref{eq:j dependent map-based sim} and \ref{eq:polarization j dependent} as another form of fast map-based simulation which retains scanning strategy structure. In this case provided the survey consists of a set of CESs we may use the constraints of NERS to capture the effects of signals which vary with crossing angle.

\subsection{Non-Circular Beams}
Until now we have dealt with circularly symmetric beams -- however in reality it is likely that the beams will be elliptical to some extent. A consequence of this is that when performing a scan the convolution of the beam with the sky will now be dependent on the crossing angle -- the orientation of the elliptical axes dictating the power that different parts of the sky contribute. This section lays out our setup and conventions, and shows why non-circular beams break the assumptions made in section \ref{section:Map-Based Sims}.

Given the dependence on crossing angle it is convenient to use spherical harmonics to describe the beam convolved TOD. We may define the relevant beams according to their spin weighted spherical harmonic transform. The response of the spin-0 temperature and spin-2 polarization beams centred at the North pole are described by
\begin{equation}
\begin{split}
    &b^{*}_{0lk} = \int d\Omega {}_{ 0}Y_{lk}^*(\Omega)B^{T}(\Omega)\text{,}
    \\
    &b^{*}_{\pm 2 lk} = \int d\Omega {}_{ \pm 2}Y_{lk}^*(\Omega)[B^{Q}(\Omega) \pm iB^{U}(\Omega)]\text{,}
\end{split}
\end{equation}
where $B^{T}(\Omega)$, $B^{Q}(\Omega)$, and $B^{U}(\Omega)$ are the beams that apply to the $I$, $Q$, and $U$ fields respectively, and ${}_{s}Y_{lk}(\Omega)$ are the spin weighted spherical harmonics.

Accordingly the detector equation describing one TOD element can be written generally for spin-$s$ fields as
\begin{equation}
t_j = \sum_{slmk}D^{l*}_{mk}(\omega_j)b^{*}_{slk}a_{slm}
\end{equation}
where $b^{*}_{slk}$, and $a_{slm}$ are the spin-$s$ weighted spherical harmonic decomposition of the beam and on-sky fields respectively whose indices $l$ and $k$, and $l$ and $m$ correspond to the multipole expansion of the beam and on-sky fields respectively. $\omega=[\phi,\theta,\psi]$ are the Euler angles describing the rotation of the beam, and $D^{l*}_{mk}$ is the Wigner-D matrix that performs the rotations on the spherical harmonic decomposition of a function \citep[e.g.][]{2014MNRAS.442.1963W}.\footnote{One can either perform the rotations clockwise (left-handed) which matches \cite{1967JMP.....8.2155G} or anti-clockwise (right-handed) \cite[e.g.][]{2000PhRvD..62l3002C,2017A&A...598A..25H}. Care is required as this changes the signs of some of the exponents.} The Euler angle rotations are performed as a rotation of the beam around the $z$ axis by $\psi$, followed by rotating by $\theta$ around the $y$ axis, followed by a rotation around the $z$ axis once more by $\phi$ -- we shall adopt the convention where these are performed in a right-handed sense.

From \cite{1967JMP.....8.2155G} (but with a change of sign as we are using the anti-clockwise rotation convention) we may write the Wigner-D matrix as
\begin{equation}
    D^{l}_{mk}(\phi,\theta,\psi) = e^{-im\phi}e^{-ik\psi} d^{l}_{mk}(\theta) = \sqrt{\frac{4\pi}{2l+1}} e^{-ik\psi} {}_{k}Y_{lm}(\theta,\phi)
\end{equation}
which we substitute to give
\begin{equation}
t_j = \sum_{slmk}\sqrt{\frac{4\pi}{2l+1}} e^{ik\psi_j} {}_{k}Y_{lm}^{*}(\theta,\phi)b^{*}_{slk}a_{slm}
\label{eq:TOD Elliptical}
\end{equation}
which describes the single polarized detector measurement of spin $s$ signals convolved with an arbitrary beam $b^{*}_{slk}$.

In the previous sections we made a simplification in equation \ref{eq:RHS Simplification} by assuming circular beams, which essentially removes the $\psi$ dependence of the leaked signal itself. However when considering an arbitrary beam as in equation \ref{eq:TOD Elliptical} we should note that the $k$-dependence of the beam means it can not be decoupled from the scan term, $e^{ik\psi_j}$. Consequently, the leaked signal (as well as the scanning terms) will now also vary with $\psi_j$.\footnote{It has been noted, by e.g. \cite{PhysRevD.77.083003}, that the effect of beam ellipticity can be well approximated by using the second spatial derivatives of $I$, $Q$, and $U$. This is similar to our use of the first spatial derivative for the pointing systematic. We shall explore in future work the possibility of this making the $\tilde{h}_n$ approach of equation \ref{eq:3x3 hn map making final} viable for beam ellipticity effects.}

The $k$-dependence of the non-circular beam smearing the signal means that the observed signal no longer has a pure spin dependence. This is different to the systematics we examined earlier as they were smoothed by a circular beam which does not have $k$-dependence.

\subsection{Application of the CES approach to Differential Ellipticity}
The polarized detectors observe a signal with contributions from the $s=0$ temperature signal and the $s=\pm2$ polarization signals. We may write the observed signal as
\small
\begin{equation}
\begin{split}
t_j &= \sum_{lmk}\sqrt{\frac{4\pi}{2l+1}} e^{ik\psi_j} {}_{k}Y_{lm}^{*}(\theta,\phi) \left( b^{*}_{0lk}a_{0lm} +b^{*}_{-2lk}a_{-2lm} + b^{*}_{2lk}a_{2lm} \right).
\end{split}
\end{equation}
\normalsize

In the case of a pair differencing experiment following equation \ref{eq:differencing} we may write
\small
\begin{equation}
\begin{split}
S &= \sum_{lmk}\sqrt{\frac{4\pi}{2l+1}} e^{ik\psi_j} {}_{k}Y_{lm}^{*}(\theta,\phi) \big((b^{A*}_{0lk} - b^{B*}_{0lk})a_{0lm}
\\
&+ (b^{A*}_{-2lk} - b^{B*}_{-2lk})a_{-2lm} + (b^{A*}_{2lk} - b^{B*}_{2lk})a_{2lm} \big)\text{.}
\end{split}
\end{equation}
\normalsize
Any differences between the $s=0$ beams of detector A and B within a pair will result in systematic leakage of the temperature to polarization and differences between the $s=\pm2$ beams will cause polarization mixing. The dominant source is the $T \rightarrow P$ leakage \citep[e.g.][]{2014MNRAS.442.1963W}.

When performing the TOD simulations it is possible to perform the beam convolution and rotation according to equation \ref{eq:TOD Elliptical} exactly in $a_{lm}$ space. However this is a time costly process which doesn't scale well to larger surveys. Instead, we follow a similar approach to the FEBeCoP code of \cite{2011ApJS..193....5M} which applies fast effective beam convolution in the pixel domain. However, unlike FEBeCoP we do not use effective beams but rather apply the beam convolution for each TOD element individually \citep{2011ApJS..193....5M,2014MNRAS.442.1963W}.

The elliptical beam we use in our simulations is given by the standard equation for an elliptical Gaussian beam. In the $x,y$ plane this is
\begin{equation}
    B(x,y) = \frac{1}{2\pi \sigma_{x}\sigma_{y}}e^{\frac{-x^{2}}{2\sigma_{x}^{2}} - \frac{y^{2}}{2\sigma_{y}^{2}}}\text{,}
\end{equation}
where $\sigma_{x}$ and $\sigma_{y}$ are the semi-major and semi-minor axes of the ellipse. This may be written in polar coordinates as
\begin{equation}
    B(r,\beta) = \frac{1}{2\pi q\sigma^2}e^{\frac{-r^2}{2\sigma^2}\left(\cos^{2}(\beta) + q^{-2} \sin^{2}(\beta)\right)}\text{,}
    \label{eq:Elliptical Beam}
\end{equation}
where $x=r\cos(\beta)$ and $y=r\sin(\beta)$, $q=\sigma_x / \sigma_y$ and $\sigma \equiv \sigma_x$ \citep{2015MNRAS.453.2058W}. The degree of asymmetry of the beam is thus defined by the parameter $q$ with the full with at half maximum (FWHM) given by $2.35\sigma$. In the case where $q=1$ the beam is axisymmetric, i.e. a circular Gaussian -- the beam smeared signal would no longer depend on the orientation angle in this scenario.

\subsubsection{Ellipticity TOD Setup}
\label{section:EllipTOD}
We perform TOD simulations as detailed in section \ref{section:Full TOD Simulation}, with the slight modification that we now convolve by an elliptical beam of FWHM $30'$ instead of circular. We will compare the results to a simulation run with a circular beam of FWHM $30'$. We use the ``Shallow'' scan strategy as detailed in table \ref{tab:scans}.

In pixel space the convolution is given as an integral over all space as
\begin{equation}
\begin{split}
    t_j = \int d\Omega &\Big(B^T_j(\Omega)T(\Omega) + B^Q_j(\Omega)Q(\Omega)\cos(2\psi_j)
    \\
    &+ B^U_j(\Omega)U(\Omega)\sin(2\psi_j)\Big)
    \label{eq:convolution}
\end{split}
\end{equation}
where we have absorbed the rotation to the correct orientation angle into the beam terms here. We assume a perfectly copolar response of the detectors which simplifies the convolution of the polarized signals making them equivalent to the temperature case. This would otherwise be complicated by cross-polar leakage requiring a polarization efficiency factor to be included \citep{2011ApJS..193....5M}. It is only the convolution procedure that is affected by this choice, the speed up introduced by the CES map-based simulation method would still remain even when including the cross-polar leakage.

We approximate the convolution integral of equation \ref{eq:convolution} in pixel space using a brute force sum. We perform the sum only on a subset of pixels within $5\sigma q$ of the pointing centre of the beam. This is a good approximation as the majority of the power of the beam is contained within that region. We verified that this choice gave us consistent results by comparing to the full convolution in $a_{lm}$ space using equation \ref{eq:TOD Elliptical} for a subset of data.

Input maps for the simulation are generated by convolving the initial unsmoothed $I$, $Q$, and $U$ maps with an elliptical beam with the $x$-axis orientated along $\psi \in \{0,\pi\}$ in 20 steps. We make use of the symmetry of the beam, which dictates that the convolved signals with orientation $\psi$ are the same as those at $(\psi+\pi)$, to create beam-convolved maps at beam orientations between $0$ and $2\pi$. These can then be linearly interpolated along the $\psi$ dimension to calculate the TOD for each pointing direction for arbitrary $\psi$ due to the functions having smooth $\psi$ dependence.

For our demonstration we set the asymmetry parameter $q=1.05$. We set the elliptical axis such that it is aligned with the detector polarization angle for the $A$ detectors within a pair, and we set the elliptical axis to be rotated $\pi/4$ with respect to the detector angles for the $B$ detectors within a pair. This can be easily varied, and we see similar agreement between the TOD and map-based results for other orientations of the elliptical axis. Both $q$ and the elliptical axis can be changed without affecting the speed up of the CES map-based simulation approach.

\subsubsection{TOD vs Map-based comparison}
\label{section:EllipTODvsMBS}
\begin{figure*}
  \centering
  \includegraphics[width=2\columnwidth]{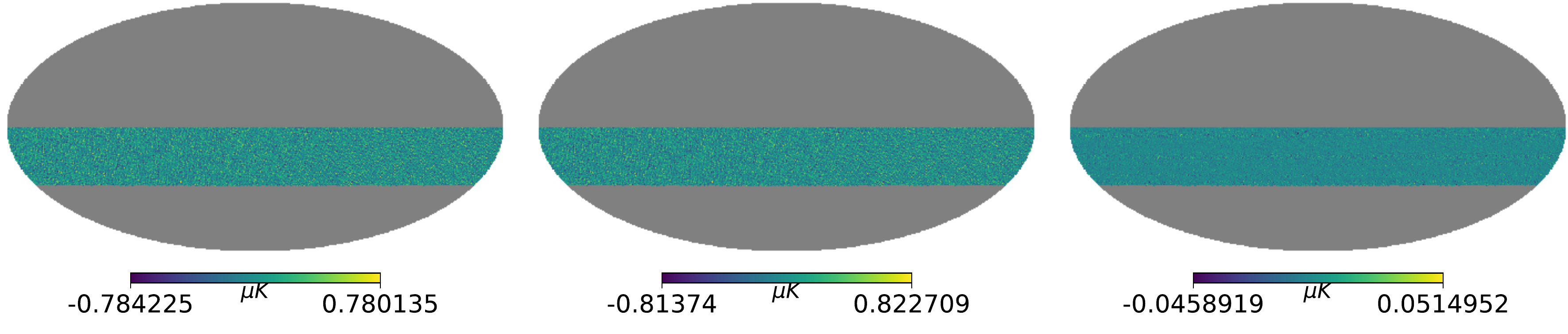}
  \includegraphics[width=2\columnwidth]{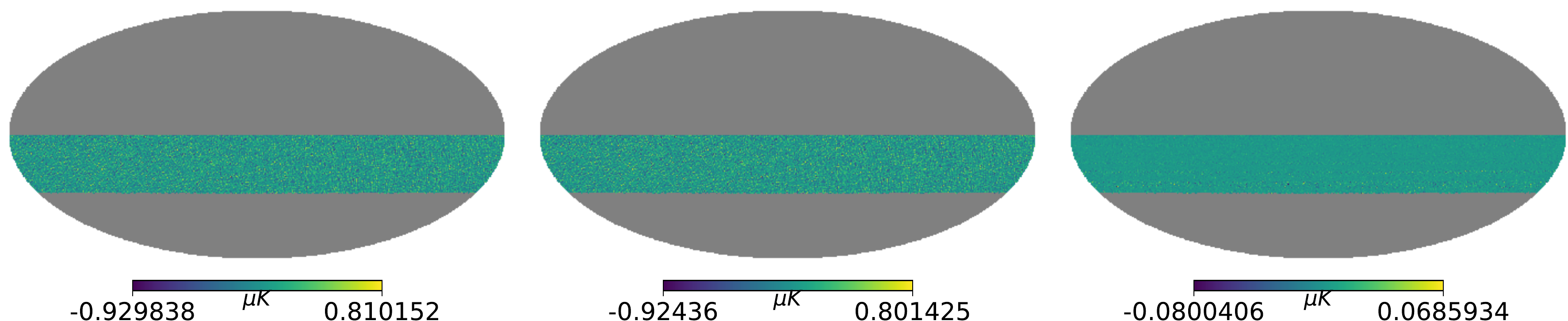}
\caption{Differential ellipticity systematic field - created by taking the difference between the output of a simulation containing the systematic and the no systematic map (the no-systematic simulation is run with a circular beam of FWHM $30'$). Top row - Q. Bottom row - U. Left column - TOD simulation using the ``Shallow'' ground-based scan. Middle column - CES map-based simulation using the ``Shallow'' scan. Right column - Residuals between the first two columns. The small level of residual indicates good agreement between the two methods. The map-based approach captures the features of the full TOD simulation well.}
\label{figure:ShallowEllipticityMaps}
\end{figure*}

Figure \ref{figure:ShallowEllipticityMaps} shows the output of the TOD simulation including the elliptical beam compared to the CES map-based approach where we have used the ``Shallow'' ground-based scan strategy. We plot the contribution of the systematic to the polarization maps by taking the difference between the output of the simulations including the differential ellipticity systematic and the no systematic map which has been generated using a circular beam of FWHM $30'$. We see good agreement between the two approaches as is evidenced by the small residual. The CES map-based approach captures the features of the full TOD simulation well.

\begin{figure}
  \centering
  \includegraphics[width=\columnwidth]{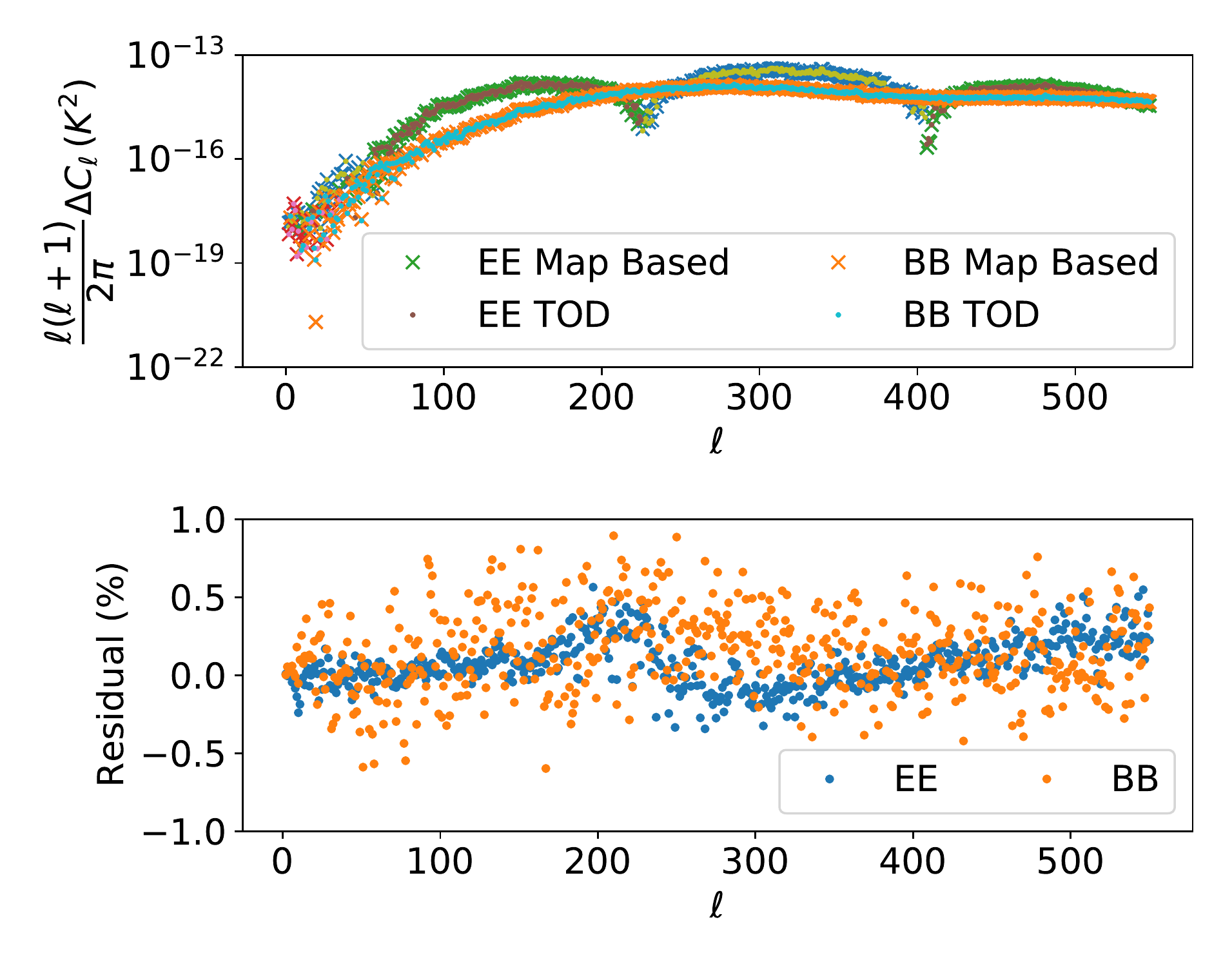}
\caption{Power spectrum of the differential ellipticity systematic. $\Delta C_{\ell}$ indicates it is the difference between a simulation containing the systematic and a simulation with no systematic present leaving just the systematic leaked signal. Upper panel: The green(/blue) crosses indicate where the $E$-mode map-based data is positive(/negative). The brown(/yellow) dots indicate where the $E$-mode TOD data is positive(/negative). The orange(/red) crosses indicate where the $B$-mode map-based data is positive(/negative). The cyan(/pink) dots indicate where the $B$-mode TOD data is positive(/negative). The ``Shallow'' ground-based TOD and map-based simulation match well. Lower Panel: The residuals are sub-percent level for the entire range of interest, showing the CES map-based approach matches the full TOD well.}
\label{figure:ShallowEllipCls}
\end{figure}

Figure \ref{figure:ShallowEllipCls} shows the resulting pseudo power spectra contribution due to the differential ellipticity systematic for both the TOD simulation and the CES map-based approach. In this case we plot $\Delta C_{\ell}$ which is the difference between a simulation containing the systematic and a simulation with no systematic present -- this isolates the systematic contribution from the true signal. We see good agreement between the full TOD and CES map-based simulation as is evidenced by the small residual which shows the approach is accurate to sub-percent level.


We further note that the structure present is dominated by temperature leakage into the $B$-mode spectrum due to the quadrupole nature of differential ellipticity leaking according to a scaled $\ell^{4} C_{\ell}^{TT}$ term. The leakage to the $E$-mode is further complicated by the cross term between the systematic and on sky signal leading to the dominant term being a scaled $\ell^{2} C_{\ell}^{TE}$ \citep{2020arXiv200800011M}. See Appendix \ref{section:Beam Structure} for more information.

This approach does not perform quite as well as the previous method presented in section \ref{section:Map-Based Sims}. However it does still achieve sub-percent level agreement with the full TOD simulation. Furthermore this alternative method has the added benefit of dealing with systematic signals that vary with crossing angle.

For our 30 day ``Shallow'' survey example, the full TOD simulation takes $\sim 9$ minutes to run, whereas the CES map-based approach takes $\sim 3$ seconds. Utilising the CES map-based approach offered a speed up of $\sim 180$ times for this particular survey. We thus have another method that offers a significant speed-up compared to a full TOD simulation while still delivering accurate simulations of systematics, including the effects of the scanning strategy.

\section{Full Focal Plane -- Correlated Systematics}
\label{section:Full Focal Plane}
In the case of a full focal plane one often makes maps for each timestream individually as
\begin{equation}
    \hat{D}_e = M^{-1}_e D_e
    \label{eq:DeMapMake}
\end{equation}
where each focal plane element $e$ has an associated vector $\hat{D}_e$ whose length corresponds to the number of fields to be measured e.g. solving for temperature and polarization would require a 3 component vector. $M_e$ is the map-making matrix containing the pointing information made up of various $\tilde{h}_{n,e}$ quantities related to the timestream, and $D_e$ is a vector of the observed signals from a given timestream. The subscript $e$ denotes the element in a focal plane we are discussing i.e. what is contributing the TOD signal -- this could be a single detector timestream, or a differenced or summed timestream from a detector pair, or the combined timestream of two detector pairs oriented as ``$+\times$'' etc. $e$ tells us which of the $N_{\text{element}}$ focal plane elements we are considering. Each of these would have their own $\tilde{h}_{n,e}$ due to being located in a different position on the focal plane. We note for context that, when applying simple binned map-making to solve for spin-0 and spin-$\pm2$ signals simultaneously, equation \ref{eq:DeMapMake} corresponds to equations \ref{eq:3x3 map making} and \ref{eq:3x3 hn map making} for a given focal plane element $e$.

Each component of the $D_e$ vector is populated according to equation \ref{eq:spinsyst} for the spin $k$ signals we wish to reconstruct as
\begin{equation}
    {}_{k}\tilde{S}^{d}_e(\Omega) = \sum_{k'=-\infty}^{\infty} \tilde{h}_{k-k',e}(\Omega){}_{k'}\tilde{S}_{e}(\Omega) \text{,}
\end{equation}
where each $\tilde{h}_{k-k',e}(\Omega)$ will differ based on the detector angle of the focal plane element $e$ measured with respect to the crossing angle and its position on the focal plane. The appropriate map-making matrix $M_e$ must then be applied in order to obtain the measured $\hat{D}_e$ signal.

Final co-added maps for a full focal plane can then be estimated by averaging (potentially using a weighting scheme) the measured per-element maps as
\begin{equation}
    \hat{D} = \frac{1}{N_\text{element}}\sum_e^{N_\text{element}} \hat{D}_e.
    \label{eq:FullFoc}
\end{equation}
However some systematics affecting the detectors will be correlated to some degree across the focal plane. How correlated the systematics are between detectors will depend on what is sourcing the systematic. Whether it is due to e.g. local or global temperature fluctuations in the instrument, steps in the fabrication process, or stems from some other issue will make a difference to how correlated we expect them to be.

In this section, we show how our approach is easily and quickly generalised to a full focal plane, including systematics that are correlated between focal plane elements.

\subsection{Correlation Method}
\label{section:Correlation Method}
In order to simulate correlation of the systematics between detectors we proceed as follows \citep{brown2009}:

For the $N_\text{detector}$ detectors we generate an $N_\text{detector} \times N_\text{detector}$ correlation matrix $C_{d,d'}$ where $C_{d,d'}=1$ for $d=d'$. The off diagonals are set to some value $0 \leq C_{d,d' \neq d} \leq 1$ where the value chosen indicates the degree of correlation i.e.
\begin{itemize}
    \item $C_{d,d' \neq d}=1$ indicates 100\% correlation.
    \item $C_{d,d' \neq d}=0.5$ indicates 50\% correlation.
    \item $C_{d,d' \neq d}=0.0$ indicates 0\% correlation.
\end{itemize}

We start with $N_\text{detector}$ independent systematics $\mathcal{S}_e$ ($\mathcal{S}_e$ could for example be the gain levels, the pointing offsets, or the level of beam asymmetry of the detectors etc.) which are initially uncorrelated having been randomly sampled from a normal distribution centred at 0 with a scatter selected according to a sensible systematic level. In order to inject the correlation we desire we first take the Cholesky decomposition of the correlation matrix $L_{d,d'}$ defined according to
\begin{equation}
    C_{d,d'} = \sum_{d''}L_{d,d''}L_{d',d''},
\end{equation}
which we then apply to $\mathcal{S}_e$ as
\begin{equation}
    \mathcal{S}^{\text{correlated}}_e = \sum_{d'}L_{d,d'}\mathcal{S}^{\text{uncorrelated}}_{d'}
\end{equation}
to generate a set of correlated systematics across the focal plane.

\subsection{Simulation}
We now present full focal plane simulations with the total number of detectors set to $1000$: we choose a focal plane setup with $250$ $+$ oriented pairs and $250$ $\times$ oriented pairs. In this case, since we still use a ``$+\times$'' focal plane element setup, we once again apply the simplifications of section \ref{section:+XSimplification}. For simplicity, we have treated all detectors as colocated on the focal plane (so that they all have the same $\tilde{h}_n$ maps). However this is not a requirement of the fast map-based simulation approach which would work just as well in the case where each focal plane element has a different location on the focal plane and hence a (slightly) different set of $\tilde{h}_n$ maps.

We perform a demonstration of this process using a differential gain systematic and thus combine equations \ref{eq:FullFoc} and \ref{eq:Map Based Differential Gain} to write the measured polarization signal by averaging the maps made across the focal plane as
\begin{equation}
\begin{split}
    &(\hat{Q}+i\hat{U})(\Omega)
    \\
    &= \frac{1}{N_\text{element}} \sum_{e}^{N_\text{element}} \big[(Q+iU)_e(\Omega) + \frac{1}{2} \tilde{h}_{2,e}(\Omega) (\delta g_{1,e} - i\delta g_{2,e}) \, I_e(\Omega)
    \\
    &+ \frac{1}{4}\tilde{h}_{0,e}(\Omega)(g^A_{1,e} + g^B_{1,e} + g^A_{2,e} + g^B_{2,e}) \, (Q+iU)_{e}(\Omega)
    \\
    &+ \frac{1}{4}\tilde{h}_{4,e}(\Omega)(g^A_{1,e} + g^B_{1,e} - g^A_{2,e} - g^B_{2,e}) \, (Q-iU)_{e}(\Omega)\big] \text{,}
\end{split}
\end{equation}
where $N_\text{element} = 250$ in this case due to the fact we are treating the combined streams of a $+$ pair and $\times$ pair together as one focal plane element.

For the gain systematic we generate two initial distributions $\mathcal{S}_e$ using different seeds with a scatter of $0.01$ indicative of a 1\% gain offset. One distribution applies to the gain $g_i^A$ of the $N_\text{detector}=500$ $A$ detectors, and the other to $g_i^B$ of the $N_\text{detector}=500$ $B$ detectors. We then apply the process of section \ref{section:Correlation Method} to apply various levels of correlation of the systematic across the focal plane between the detectors -- we do this independently for the 500 $A$ and 500 $B$ detectors. We examine the cases of complete correlation $C_{e,e' \neq e}=1$, completely uncorrelated $C_{e,e' \neq e}=0.0$, and an intermediate case of $C_{e,e' \neq e}=0.5$.

We note that the choice of how to correlate the detectors here is arbitrary. However, it serves the purpose of illustrating that the fast map-based simulation approach can readily and accurately approximate  maps resulting from the coaddition of large numbers of detectors from an extended focal plane. If a given experiment had access to better estimates of the expected systematic levels of each of its detectors in the focal plane this could trivially be incorporated into the method we present.

\begin{figure}
  \centering
  \includegraphics[width=\columnwidth]{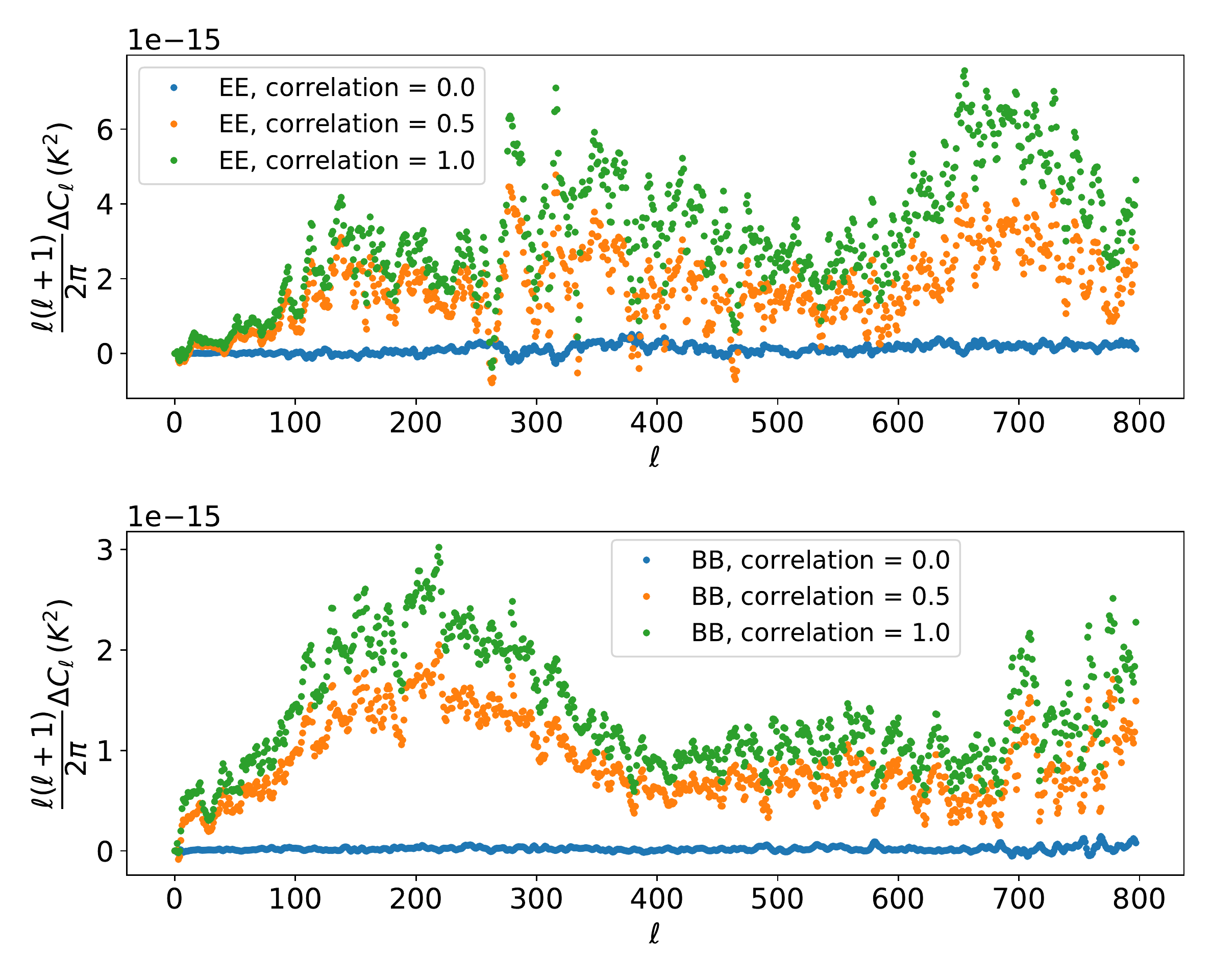}
\caption{The power spectrum of the differential gain systematic applied when averaging across a focal plane of 1000 detectors, arranged as 250 $+$ oriented and 250 $\times$ oriented pairs -- $\Delta C_{\ell}$ indicates it is the difference between a simulation containing the systematic and a simulation with no systematic present leaving just the systematic leaked signal. Upper Panel - $E$-mode. Lower Panel - $B$-mode. Green points show the case of completely correlated systematics across detectors, orange points show the 50\% correlated case, and blue points show the uncorrelated case. We see that in the case where the systematics are uncorrelated across the focal plane the result of averaging the maps leads to the gain systematic being suppressed -- the power spectrum losing the structure of the leakage and essentially becoming a small scatter about 0. In the fully correlated case the systematics do not average down and we see a large systematic signal dominated by temperature leakage in the $B$-mode, whilst in the $E$-mode the temperature leakage dominates at low-$\ell$ and the amplification of the polarization signal dominates at $\ell \gtrsim 300$. The leaked signal in the intermediate case of 50\% correlation has been suppressed slightly, by the process of averaging the observed maps, compared to the fully correlated case.}
\label{figure:fullfocgaincls}
\end{figure}

Figure \ref{figure:fullfocgaincls} shows the output pseudo power spectra generated by applying the fast map-based simulation to a full focal plane of 1000 detectors for the ``Deep'' survey. We once again plot $\Delta C_{\ell}$ which is the difference between the simulations containing the systematics and simulations with no systematic present -- this isolates the systematic contribution from the true signal. We see as expected that if systematics are uncorrelated between detectors across the full focal plane the systematic signal will average down, resulting in a noisy scatter close to zero, for both the $E$- and $B$-mode.

In the completely correlated case the systematic signal has clear structure from the leaked temperature and polarization mixing. The systematic has not averaged down due to being highly correlated between all detectors. The dominant signal is the temperature leakage in the case of the $B$-mode, while in the case of the $E$-mode we see the temperature leakage dominates at low-$\ell$ before the amplification of the polarization signal begins to dominate at higher-$\ell$ causing the characteristic peaks at $\ell \approx 400$ and $\ell \approx 700$. Further detail on this is available in \cite{2020arXiv200800011M}.

In the intermediate case of $50\%$ correlation between detectors the leakage has been suppressed more than the fully correlated case due to the averaging process. However, it still retains structure from the leaked systematic signals. The process essentially results in a continuum in which the less correlated the detectors the better the systematic will average down when averaging maps from the full focal plane.

The map-based approach will enable future studies of correlated systematic effects across the focal plane to be performed rapidly for many detectors. In the uncorrelated case it will aid in characterising how well the systematic is suppressed and the expected levels of scatter on the data resulting from averaging measurements from the full focal plane. In the correlated case it will aid in identifying the expected levels of systematics and the footprint they will leave in the data for different degrees of correlation.

More importantly we have shown how to apply the fast map-based simulation to a full focal plane simulation. A full TOD simulation would be very time costly when simulating a full focal plane. However the map-based simulation for the ``Deep'' survey ran for the 1000 detector focal plane in less than 1 minute. Provided we have access to the required $\tilde{h}_n$ scan maps of a survey, for the detectors in the focal plane, we can speedily create simulations of a number of systematics whilst retaining effects of the scanning strategy.

\section{Conclusions}
\label{section:Conclusion}
We have presented two techniques that utilise summary properties of the scanning strategy to rapidly simulate systematics in CMB surveys whilst retaining structure from the scanning strategy. For many applications, these approaches remove the need to repeatedly run computationally expensive full TOD simulations. Using the stored scanning data, from a single run of a TOD simulation, will facilitate rapid forecasts of the effects of a number of systematics on upcoming CMB surveys. The map-based approach can incorporate varying magnitudes of systematics, differing CMB realisations, and realistic numbers (thousands) of detectors which means it can be used in tandem with monte-carlo techniques without incurring the significant computational expense of full TOD simulations.

The key results are as follows:
\begin{itemize}
    \item We first presented a fast map-based approach which utilises summary properties of a scanning strategy to rapidly perform simulations of systematics in CMB surveys for both ground-based, and space-based instruments. By using stored $\tilde{h}_n$ maps generated from the scanning strategy we are able to shortcut the TOD process for the case of time-independent systematics by performing only a single calculation per pixel in map-space -- the key equations are equation \ref{eq:spinsyst} which shows how to calculate the leakage of fields of arbitrary spin, and equations \ref{eq:RHS Simplification} and \ref{eq:3x3 hn map making final} which show how these quantities relate to map-making in the CMB context. We showed that, by sacrificing some of the complexities of a full TOD simulation, one can utilise this map-based process to simulate systematics, whilst retaining structure from the scanning strategy, in a fraction of the time of a full TOD simulation. We demonstrated this for both a differential gain and differential pointing systematic. The TOD simulation of a 1-year scan by the proposed {\it EPIC} satellite took 10--15 hours of CPU time whilst the associated map-based simulation took $\sim$10 seconds, i.e. the map-based simulation was $\sim$3600--5400 times faster. However the scan parameters are not the limiting factor for the map-based approach so the speed up for longer TOD sets would be even greater. 
    
    \item The second approach we presented utilises stringent constraints that are placed upon the crossing angle coverage for ground-based instruments when implementing constant elevations scans. By exploiting these constraints we can shortcut the TOD process by using stored maps of the allowed crossing angles for each scanning elevation. Combining these with the corresponding hit maps generated separately for when the target observation field is rising and setting, we can utilise equation \ref{eq:j dependent map-based sim} to fast track the simulation of systematic effects on a CMB survey. The 30 day ``Shallow'' TOD simulation we used took $\sim 9$ minutes of CPU time whereas the corresponding map-based simulation took $\sim 3$ seconds, i.e. the CES map-based approach offered a speed up of $\sim 180$ times for this particular survey. Once again, we note that the scan parameters are not the limiting factor for the CES map-based approach so we would see greater speed ups for larger TOD simulations. This CES approach does not perform quite as well as the $\tilde{h}_n$ approach but it does offer an additional benefit: it is able to accurately model systematic effects that vary with crossing angle. There are a number of systematics this could be useful for as we demonstrated with the differential ellipticity systematic. Furthermore despite a slight dip in performance compared to the first method, this approach does still offer sub-percent agreement with the full TOD simulation and still offers a significant speed up while incorporating effects of the scanning strategy.
    
    \item Finally we showed how to extend the fast map-based simulation method to process a full focal plane. Full TOD simulations are extremely computationally expensive especially when scaling up to a full focal plane -- the map-based approach offers an alternative method to simulate systematics which is far less time intensive and still includes structure from the scanning strategy. We demonstrated how this can be implemented using a differential gain systematic, showing that the fast map-based approach offers a way to consider the effects of systematics across a full focal plane rapidly.
\end{itemize}

The techniques presented in this paper will be useful for speeding up forecasting and tolerancing studies of systematic effects for the next generation of ground-based and satellite CMB surveys, leading up to and including “Stage IV” projects. Our methods could also be useful other forthcoming cosmological surveys that employ scanning observations, in particular intensity mapping surveys.

\section*{Acknowledgements}
NM is supported by a STFC studentship. DBT acknowledges support from Science and Technology Facilities Council (STFC) grants ST/P000649/1, ST/T000414/1 and ST/T000341/1.

\section*{Data Availability}
The algorithms we employ here and demonstration scripts for reproducing the results of this article are available at \url{https://github.com/NiumCosmo/MBSS4CMB}



\bibliographystyle{mnras}
\bibliography{Map_Based_Sims} 

\begin{thebibliography}{}
\makeatletter
\relax
\def\mn@urlcharsother{\let\do\@makeother \do\$\do\&\do\#\do\^\do\_\do\%\do\~}
\def\mn@doi{\begingroup\mn@urlcharsother \@ifnextchar [ {\mn@doi@}
  {\mn@doi@[]}}
\def\mn@doi@[#1]#2{\def\@tempa{#1}\ifx\@tempa\@empty \href
  {http://dx.doi.org/#2} {doi:#2}\else \href {http://dx.doi.org/#2} {#1}\fi
  \endgroup}
\def\mn@eprint#1#2{\mn@eprint@#1:#2::\@nil}
\def\mn@eprint@arXiv#1{\href {http://arxiv.org/abs/#1} {{\tt arXiv:#1}}}
\def\mn@eprint@dblp#1{\href {http://dblp.uni-trier.de/rec/bibtex/#1.xml}
  {dblp:#1}}
\def\mn@eprint@#1:#2:#3:#4\@nil{\def\@tempa {#1}\def\@tempb {#2}\def\@tempc
  {#3}\ifx \@tempc \@empty \let \@tempc \@tempb \let \@tempb \@tempa \fi \ifx
  \@tempb \@empty \def\@tempb {arXiv}\fi \@ifundefined
  {mn@eprint@\@tempb}{\@tempb:\@tempc}{\expandafter \expandafter \csname
  mn@eprint@\@tempb\endcsname \expandafter{\@tempc}}}

\bibitem[\protect\citeauthoryear{{Abazajian} et~al.,}{{Abazajian}
  et~al.}{2016}]{2016arXiv161002743A}
{Abazajian} K.~N.,  et~al., 2016, arXiv e-prints, \href
  {https://ui.adsabs.harvard.edu/abs/2016arXiv161002743A} {p. arXiv:1610.02743}

\bibitem[\protect\citeauthoryear{Ade et~al.}{Ade et~al.}{2014}]{1403.2369}
Ade P. A.~R.,  et~al., 2014, \mn@doi [Astrophys. J.]
  {10.1088/0004-637X/794/2/171}, 794, 171

\bibitem[\protect\citeauthoryear{{Ade} et~al.,}{{Ade}
  et~al.}{2019}]{2019JCAP...02..056A}
{Ade} P.,  et~al., 2019, \mn@doi [\jcap] {10.1088/1475-7516/2019/02/056}, \href
  {https://ui.adsabs.harvard.edu/abs/2019JCAP...02..056A} {2019, 056}

\bibitem[\protect\citeauthoryear{{Bicep2 Collaboration} et~al.,}{{Bicep2
  Collaboration} et~al.}{2015}]{2015ApJ...814..110B}
{Bicep2 Collaboration} et~al., 2015, \mn@doi [\apj]
  {10.1088/0004-637X/814/2/110}, \href
  {https://ui.adsabs.harvard.edu/abs/2015ApJ...814..110B} {814, 110}

\bibitem[\protect\citeauthoryear{{Blas}, {Lesgourgues}  \& {Tram}}{{Blas}
  et~al.}{2011}]{2011JCAP...07..034B}
{Blas} D.,  {Lesgourgues} J.,   {Tram} T.,  2011, \mn@doi [\jcap]
  {10.1088/1475-7516/2011/07/034}, \href
  {https://ui.adsabs.harvard.edu/abs/2011JCAP...07..034B} {2011, 034}

\bibitem[\protect\citeauthoryear{{Bock} et~al.,}{{Bock}
  et~al.}{2008}]{2008arXiv0805.4207B}
{Bock} J.,  et~al., 2008, arXiv e-prints, \href
  {https://ui.adsabs.harvard.edu/abs/2008arXiv0805.4207B} {p. arXiv:0805.4207}

\bibitem[\protect\citeauthoryear{{Bock} et~al.,}{{Bock}
  et~al.}{2009}]{2009arXiv0906.1188B}
{Bock} J.,  et~al., 2009, arXiv e-prints, \href
  {https://ui.adsabs.harvard.edu/abs/2009arXiv0906.1188B} {p. arXiv:0906.1188}

\bibitem[\protect\citeauthoryear{{Bowden} et~al.,}{{Bowden}
  et~al.}{2004}]{2004MNRAS.349..321B}
{Bowden} M.,  et~al., 2004, \mn@doi [\mnras]
  {10.1111/j.1365-2966.2004.07506.x}, \href
  {http://adsabs.harvard.edu/abs/2004MNRAS.349..321B} {349, 321}

\bibitem[\protect\citeauthoryear{{Brown}, {Challinor}, {North}, {Johnson},
  {O'Dea}  \& {Sutton}}{{Brown} et~al.}{2009}]{brown2009}
{Brown} M.~L.,  {Challinor} A.,  {North} C.~E.,  {Johnson} B.~R.,  {O'Dea} D.,
   {Sutton} D.,  2009, \mn@doi [\mnras] {10.1111/j.1365-2966.2009.14975.x},
  \href {http://adsabs.harvard.edu/abs/2009MNRAS.397..634B} {397, 634}

\bibitem[\protect\citeauthoryear{{Challinor}, {Fosalba}, {Mortlock}, {Ashdown},
  {Wandelt}  \& {G{\'o}rski}}{{Challinor} et~al.}{2000}]{2000PhRvD..62l3002C}
{Challinor} A.,  {Fosalba} P.,  {Mortlock} D.,  {Ashdown} M.,  {Wandelt} B.,
  {G{\'o}rski} K.,  2000, \mn@doi [\prd] {10.1103/PhysRevD.62.123002}, \href
  {https://ui.adsabs.harvard.edu/abs/2000PhRvD..62l3002C} {62, 123002}

\bibitem[\protect\citeauthoryear{{Crowley} et~al.,}{{Crowley}
  et~al.}{2018}]{2018SPIE10708E..3ZC}
{Crowley} K.~T.,  et~al., 2018, in Millimeter, Submillimeter, and Far-Infrared
  Detectors and Instrumentation for Astronomy IX. p. 107083Z (\mn@eprint
  {arXiv} {1808.10491}), \mn@doi{10.1117/12.2313414}

\bibitem[\protect\citeauthoryear{{D'Alessandro}, {Mele}, {Columbro}, {Pagano},
  {Piacentini}, {de Bernardis}  \& {Masi}}{{D'Alessandro}
  et~al.}{2019}]{2019A&A...627A.160D}
{D'Alessandro} G.,  {Mele} L.,  {Columbro} F.,  {Pagano} L.,  {Piacentini} F.,
  {de Bernardis} P.,   {Masi} S.,  2019, \mn@doi [\aap]
  {10.1051/0004-6361/201834495}, \href
  {https://ui.adsabs.harvard.edu/abs/2019A&A...627A.160D} {627, A160}

\bibitem[\protect\citeauthoryear{{Dickinson}}{{Dickinson}}{2016}]{2016arXiv160603606D}
{Dickinson} C.,  2016, arXiv e-prints, \href
  {https://ui.adsabs.harvard.edu/abs/2016arXiv160603606D} {p. arXiv:1606.03606}

\bibitem[\protect\citeauthoryear{{Duivenvoorden}, {Gudmundsson}  \&
  {Rahlin}}{{Duivenvoorden} et~al.}{2019}]{2019MNRAS.486.5448D}
{Duivenvoorden} A.~J.,  {Gudmundsson} J.~E.,   {Rahlin} A.~S.,  2019, \mn@doi
  [\mnras] {10.1093/mnras/stz1143}, \href
  {https://ui.adsabs.harvard.edu/abs/2019MNRAS.486.5448D} {486, 5448}

\bibitem[\protect\citeauthoryear{{Durrer}}{{Durrer}}{2015}]{durrer2015}
{Durrer} R.,  2015, \mn@doi [Classical and Quantum Gravity]
  {10.1088/0264-9381/32/12/124007}, \href
  {https://ui.adsabs.harvard.edu/abs/2015CQGra..32l4007D} {32, 124007}

\bibitem[\protect\citeauthoryear{{Fa{\'u}ndez} et~al.,}{{Fa{\'u}ndez}
  et~al.}{2020}]{2020ApJ...893...85F}
{Fa{\'u}ndez} M.~A.,  et~al., 2020, \mn@doi [\apj] {10.3847/1538-4357/ab7e29},
  \href {https://ui.adsabs.harvard.edu/abs/2020ApJ...893...85F} {893, 85}

\bibitem[\protect\citeauthoryear{{Flux{\'a}}, {Brewer}  \&
  {D{\"u}nner}}{{Flux{\'a}} et~al.}{2020}]{2020JCAP...02..030F}
{Flux{\'a}} P.,  {Brewer} M.~K.,   {D{\"u}nner} R.,  2020, \mn@doi [\jcap]
  {10.1088/1475-7516/2020/02/030}, \href
  {https://ui.adsabs.harvard.edu/abs/2020JCAP...02..030F} {2020, 030}

\bibitem[\protect\citeauthoryear{{Goldberg}, {Macfarlane}, {Newman}, {Rohrlich}
   \& {Sudarshan}}{{Goldberg} et~al.}{1967}]{1967JMP.....8.2155G}
{Goldberg} J.~N.,  {Macfarlane} A.~J.,  {Newman} E.~T.,  {Rohrlich} F.,
  {Sudarshan} E.~C.~G.,  1967, \mn@doi [Journal of Mathematical Physics]
  {10.1063/1.1705135}, \href
  {https://ui.adsabs.harvard.edu/abs/1967JMP.....8.2155G} {8, 2155}

\bibitem[\protect\citeauthoryear{{G{\'o}rski}, {Hivon}, {Banday}, {Wand elt},
  {Hansen}, {Reinecke}  \& {Bartelmann}}{{G{\'o}rski}
  et~al.}{2005}]{2005ApJ...622..759G}
{G{\'o}rski} K.~M.,  {Hivon} E.,  {Banday} A.~J.,  {Wand elt} B.~D.,  {Hansen}
  F.~K.,  {Reinecke} M.,   {Bartelmann} M.,  2005, \mn@doi [\apj]
  {10.1086/427976}, \href
  {https://ui.adsabs.harvard.edu/abs/2005ApJ...622..759G} {622, 759}

\bibitem[\protect\citeauthoryear{{Green}, {Meyers}  \& {van Engelen}}{{Green}
  et~al.}{2017}]{2017JCAP...12..005G}
{Green} D.,  {Meyers} J.,   {van Engelen} A.,  2017, \mn@doi [\jcap]
  {10.1088/1475-7516/2017/12/005}, \href
  {https://ui.adsabs.harvard.edu/abs/2017JCAP...12..005G} {2017, 005}

\bibitem[\protect\citeauthoryear{{Hivon}, {Mottet}  \& {Ponthieu}}{{Hivon}
  et~al.}{2017}]{2017A&A...598A..25H}
{Hivon} E.,  {Mottet} S.,   {Ponthieu} N.,  2017, \mn@doi [\aap]
  {10.1051/0004-6361/201629626}, \href
  {https://ui.adsabs.harvard.edu/abs/2017A&A...598A..25H} {598, A25}

\bibitem[\protect\citeauthoryear{{Hu}, {Hedman}  \& {Zaldarriaga}}{{Hu}
  et~al.}{2003}]{2003PhRvD..67d3004H}
{Hu} W.,  {Hedman} M.~M.,   {Zaldarriaga} M.,  2003, \mn@doi [\prd]
  {10.1103/PhysRevD.67.043004}, \href
  {https://ui.adsabs.harvard.edu/abs/2003PhRvD..67d3004H} {67, 043004}

\bibitem[\protect\citeauthoryear{{Hui} et~al.,}{{Hui}
  et~al.}{2018}]{2018SPIE10708E..07H}
{Hui} H.,  et~al., 2018, in {Zmuidzinas} J.,  {Gao} J.-R.,  eds,  Society of
  Photo-Optical Instrumentation Engineers (SPIE) Conference Series Vol. 10708,
  Millimeter, Submillimeter, and Far-Infrared Detectors and Instrumentation for
  Astronomy IX. p. 1070807 (\mn@eprint {arXiv} {1808.00568}),
  \mn@doi{10.1117/12.2311725}

\bibitem[\protect\citeauthoryear{{Jarosik} et~al.,}{{Jarosik}
  et~al.}{2011}]{2011ApJS..192...14J}
{Jarosik} N.,  et~al., 2011, \mn@doi [\apjs] {10.1088/0067-0049/192/2/14},
  \href {https://ui.adsabs.harvard.edu/abs/2011ApJS..192...14J} {192, 14}

\bibitem[\protect\citeauthoryear{{McCallum}, {Thomas}, {Brown}  \&
  {Tessore}}{{McCallum} et~al.}{2020}]{2020arXiv200800011M}
{McCallum} N.,  {Thomas} D.~B.,  {Brown} M.~L.,   {Tessore} N.,  2020, arXiv
  e-prints, \href {https://ui.adsabs.harvard.edu/abs/2020arXiv200800011M} {p.
  arXiv:2008.00011}

\bibitem[\protect\citeauthoryear{{McCallum}, {Thomas}, {Bull}  \&
  {Brown}}{{McCallum} et~al.}{2021}]{2021arXiv210708058M}
{McCallum} N.,  {Thomas} D.~B.,  {Bull} P.,   {Brown} M.~L.,  2021, arXiv
  e-prints, \href {https://ui.adsabs.harvard.edu/abs/2021arXiv210708058M} {p.
  arXiv:2107.08058}

\bibitem[\protect\citeauthoryear{Miller, Shimon  \& Keating}{Miller
  et~al.}{2009}]{0806.3096}
Miller N.~J.,  Shimon M.,   Keating B.~G.,  2009, \mn@doi [Phys. Rev.]
  {10.1103/PhysRevD.79.063008}, D79, 063008

\bibitem[\protect\citeauthoryear{{Mitra}, {Rocha}, {G{\'o}rski},
  {Huffenberger}, {Eriksen}, {Ashdown}  \& {Lawrence}}{{Mitra}
  et~al.}{2011}]{2011ApJS..193....5M}
{Mitra} S.,  {Rocha} G.,  {G{\'o}rski} K.~M.,  {Huffenberger} K.~M.,  {Eriksen}
  H.~K.,  {Ashdown} M.~A.~J.,   {Lawrence} C.~R.,  2011, \mn@doi [\apjs]
  {10.1088/0067-0049/193/1/5}, \href
  {https://ui.adsabs.harvard.edu/abs/2011ApJS..193....5M} {193, 5}

\bibitem[\protect\citeauthoryear{{N{\ae}ss} \& {Louis}}{{N{\ae}ss} \&
  {Louis}}{2013}]{2013JCAP...09..001N}
{N{\ae}ss} S.~K.,  {Louis} T.,  2013, \mn@doi [\jcap]
  {10.1088/1475-7516/2013/09/001}, \href
  {https://ui.adsabs.harvard.edu/abs/2013JCAP...09..001N} {2013, 001}

\bibitem[\protect\citeauthoryear{{Natoli} et~al.,}{{Natoli}
  et~al.}{2018}]{2018JCAP...04..022N}
{Natoli} P.,  et~al., 2018, \mn@doi [\jcap] {10.1088/1475-7516/2018/04/022},
  \href {https://ui.adsabs.harvard.edu/abs/2018JCAP...04..022N} {2018, 022}

\bibitem[\protect\citeauthoryear{{O'Dea}, {Challinor}  \& {Johnson}}{{O'Dea}
  et~al.}{2007}]{2007MNRAS.376.1767O}
{O'Dea} D.,  {Challinor} A.,   {Johnson} B.~R.,  2007, \mn@doi [\mnras]
  {10.1111/j.1365-2966.2007.11558.x}, \href
  {https://ui.adsabs.harvard.edu/abs/2007MNRAS.376.1767O} {376, 1767}

\bibitem[\protect\citeauthoryear{{Planck Collaboration} et~al.,}{{Planck
  Collaboration} et~al.}{2014}]{2014A&A...571A...7P}
{Planck Collaboration} et~al., 2014, \mn@doi [\aap]
  {10.1051/0004-6361/201321535}, \href
  {https://ui.adsabs.harvard.edu/abs/2014A&A...571A...7P} {571, A7}

\bibitem[\protect\citeauthoryear{{Planck Collaboration} et~al.,}{{Planck
  Collaboration} et~al.}{2016a}]{2016A&A...594A...7P}
{Planck Collaboration} et~al., 2016a, \mn@doi [\aap]
  {10.1051/0004-6361/201525844}, \href
  {https://ui.adsabs.harvard.edu/abs/2016A&A...594A...7P} {594, A7}

\bibitem[\protect\citeauthoryear{{Planck Collaboration} et~al.,}{{Planck
  Collaboration} et~al.}{2016b}]{2016A&A...594A..13P}
{Planck Collaboration} et~al., 2016b, \mn@doi [\aap]
  {10.1051/0004-6361/201525830}, \href
  {https://ui.adsabs.harvard.edu/abs/2016A&A...594A..13P} {594, A13}

\bibitem[\protect\citeauthoryear{{Rhodes}}{{Rhodes}}{2011}]{2011ascl.soft12014R}
{Rhodes} B.~C.,  2011, {PyEphem: Astronomical Ephemeris for Python} (\mn@eprint
  {ascl} {1112.014})

\bibitem[\protect\citeauthoryear{{Salatino} et~al.,}{{Salatino}
  et~al.}{2018}]{2018SPIE10708E..48S}
{Salatino} M.,  et~al., 2018, in \procspie. p. 1070848 (\mn@eprint {arXiv}
  {1808.07442}), \mn@doi{10.1117/12.2312993}

\bibitem[\protect\citeauthoryear{Shimon, Keating, Ponthieu  \& Hivon}{Shimon
  et~al.}{2008}]{PhysRevD.77.083003}
Shimon M.,  Keating B.,  Ponthieu N.,   Hivon E.,  2008, \mn@doi [Phys. Rev. D]
  {10.1103/PhysRevD.77.083003}, 77, 083003

\bibitem[\protect\citeauthoryear{Staggs, Dunkley  \& Page}{Staggs
  et~al.}{2018}]{staggs2018}
Staggs S.,  Dunkley J.,   Page L.,  2018, \mn@doi [Reports on Progress in
  Physics] {10.1088/1361-6633/aa94d5}, 81, 044901

\bibitem[\protect\citeauthoryear{{Stevens} et~al.,}{{Stevens}
  et~al.}{2018}]{2018SPIE10708E..41S}
{Stevens} J.~R.,  et~al., 2018, in Millimeter, Submillimeter, and Far-Infrared
  Detectors and Instrumentation for Astronomy IX. p. 1070841 (\mn@eprint
  {arXiv} {1808.05131}), \mn@doi{10.1117/12.2313898}

\bibitem[\protect\citeauthoryear{{Suzuki} et~al.,}{{Suzuki}
  et~al.}{2018}]{2018JLTP..193.1048S}
{Suzuki} A.,  et~al., 2018, \mn@doi [Journal of Low Temperature Physics]
  {10.1007/s10909-018-1947-7}, \href
  {https://ui.adsabs.harvard.edu/abs/2018JLTP..193.1048S} {193, 1048}

\bibitem[\protect\citeauthoryear{{Tegmark}}{{Tegmark}}{1997a}]{1997PhRvD..56.4514T}
{Tegmark} M.,  1997a, \mn@doi [\prd] {10.1103/PhysRevD.56.4514}, \href
  {https://ui.adsabs.harvard.edu/abs/1997PhRvD..56.4514T} {56, 4514}

\bibitem[\protect\citeauthoryear{{Tegmark}}{{Tegmark}}{1997b}]{1997ApJ...480L..87T}
{Tegmark} M.,  1997b, \mn@doi [\apjl] {10.1086/310631}, \href
  {https://ui.adsabs.harvard.edu/abs/1997ApJ...480L..87T} {480, L87}

\bibitem[\protect\citeauthoryear{{Thomas}, {McCallum}  \& {Brown}}{{Thomas}
  et~al.}{2020}]{2020MNRAS.491.1960T}
{Thomas} D.~B.,  {McCallum} N.,   {Brown} M.~L.,  2020, \mn@doi [\mnras]
  {10.1093/mnras/stz2607}, \href
  {https://ui.adsabs.harvard.edu/abs/2020MNRAS.491.1960T} {491, 1960}

\bibitem[\protect\citeauthoryear{{Thomas}, {McCallum}  \& {Brown}}{{Thomas}
  et~al.}{2021}]{2021arXiv210202284T}
{Thomas} D.~B.,  {McCallum} N.,   {Brown} M.~L.,  2021, arXiv e-prints, \href
  {https://ui.adsabs.harvard.edu/abs/2021arXiv210202284T} {p. arXiv:2102.02284}

\bibitem[\protect\citeauthoryear{{Wallis}, {Brown}, {Battye}, {Pisano}  \&
  {Lamagna}}{{Wallis} et~al.}{2014}]{2014MNRAS.442.1963W}
{Wallis} C. G.~R.,  {Brown} M.~L.,  {Battye} R.~A.,  {Pisano} G.,   {Lamagna}
  L.,  2014, \mn@doi [\mnras] {10.1093/mnras/stu856}, \href
  {https://ui.adsabs.harvard.edu/abs/2014MNRAS.442.1963W} {442, 1963}

\bibitem[\protect\citeauthoryear{{Wallis}, {Bonaldi}, {Brown}  \&
  {Battye}}{{Wallis} et~al.}{2015}]{2015MNRAS.453.2058W}
{Wallis} C. G.~R.,  {Bonaldi} A.,  {Brown} M.~L.,   {Battye} R.~A.,  2015,
  \mn@doi [\mnras] {10.1093/mnras/stv1689}, \href
  {https://ui.adsabs.harvard.edu/abs/2015MNRAS.453.2058W} {453, 2058}

\bibitem[\protect\citeauthoryear{Wallis, Brown, Battye  \& Delabrouille}{Wallis
  et~al.}{2017}]{Wallisetal2016}
Wallis C. G.~R.,  Brown M.~L.,  Battye R.~A.,   Delabrouille J.,  2017, \mn@doi
  [Mon. Not. Roy. Astron. Soc.] {10.1093/mnras/stw2577}, 466, 425

\bibitem[\protect\citeauthoryear{Zonca, Singer, Lenz, Reinecke, Rosset, Hivon
  \& Gorski}{Zonca et~al.}{2019}]{Zonca2019}
Zonca A.,  Singer L.,  Lenz D.,  Reinecke M.,  Rosset C.,  Hivon E.,   Gorski
  K.,  2019, \mn@doi [Journal of Open Source Software] {10.21105/joss.01298},
  4, 1298

\makeatother
\end{thebibliography}




\appendix

\section{Ground-Based Scanning Strategies}
\label{section:GBSS}
In order to generate the scan strategy information for the ground-based surveys we made use of the pyEphem package \citep{2011ascl.soft12014R}. We initially define an observatory using an ephem.Observer() object which is given a latitude (-22:56.396) and longitude (-67:46.816) denoting its position on Earth.

We then set a target field to observe. We use limits of Dec=$-50^{\circ}$ to $-30^{\circ}$ and R.A = $-50^{\circ}$ to $90^{\circ}$ for the ``Deep'' field, and limits of Dec=$-35^{\circ}$ to $0^{\circ}$ and R.A = $-180^{\circ}$ to $180^{\circ}$ for the ``Shallow'' field. Subsequently we select a target observation date of 2022/01/01 and the instrument elevation we wish to observe at which we keep constant at $35^{\circ}$ for our first set of scans and then $50^{\circ}$ for our second set.

We then define an ephem.FixedBody() object to act as our initial target for where we wish to begin the scan (using the field's lower bounds of Dec and R.A for the rising scan, and the field's lower bound of R.A. and upper bound of Dec for the setting scan) and use it to calculate the next time the field will be visible to the observatory for the desired elevation. The FixedBody Compute method is then used to find the required starting azimuth for the observations. The ending azimuth for each swipe can then be computed using bounds on the target fields. The azimuth bounds for each scan rising and setting at each elevation are given in table \ref{tab:scans}.

We perform the scan by swiping between the two azimuth bounds continuously for the extent of time the field is visible each day. We perform two scans each day, one as the field rises and one as it sets. The time is calculated simply as
\begin{equation}
    t = (\text{RA}^{\text{Upper}}-\text{RA}^{\text{Lower}}) \left(\frac{23.9345}{360}\right)
\end{equation}
where $\text{RA}^{\text{Upper}}$ and $\text{RA}^{\text{Lower}}$ are the bounding right ascensions of the field in degrees, and the factor of $\frac{23.9345}{360}$ gives the sky rotation rate in hours/degree.

Finally we may generate the scan data at each time step which includes the right ascension, declination, and crossing angle. Having calculated the azimuth bounds and set a constant elevation, we may use these along with the sampling frequency (10 Hz), and azimuth slew rate ($1^{\circ}$/s) to perform the calculations required.

The R.A and Dec data are generated for each time step using the radec\_of() method of pyEphem for the known azimuth and elevation. The crossing angle $\psi$ is calculated using the two argument arctangent as
\begin{equation}
    \psi = \arctan2\left(\frac{\text{Dec}_b - \text{Dec}_a}{(\text{RA}_b - \text{RA}_a) \cos{(\text{Dec}_a)}}\right),
\end{equation}
where the subscript $a$ denotes the current value and subscript $b$ denotes the value from the subsequent time step.

The process for each scan is repeated each day for both rising and setting. We generate a scanning strategy for 30 days at each elevation.

\section{Beam Structure}
\label{section:Beam Structure}
There are many types of beam systematics which matter for CMB instruments. Whilst a perfect circular Gaussian beam would be ideal, in reality there will always be imperfections.

Differencing of orthogonal detector pairs is a commonly used technique to separate the temperature and polarization. However any imperfections between the detectors can result in the leakage of temperature to polarization and polarization mixing. This is of course a very dangerous contaminant to $B$-mode searches given the relative sizes of the signals.

\begin{figure}
  \centering
  \includegraphics[width=\columnwidth]{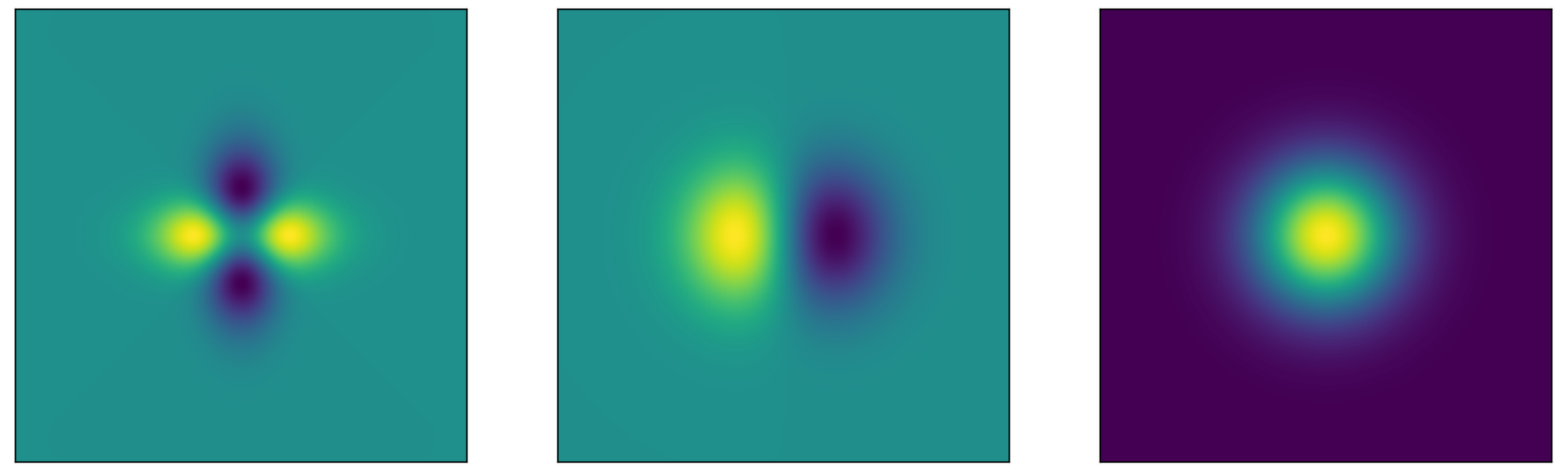}
\caption{The beam structure post detector differencing. Left panel: Differential ellipticity $\rightarrow$ quadrupole leakage. Middle panel: Differential pointing $\rightarrow$ dipole leakage. Right panel: Differential gain $\rightarrow$ monopole leakage.}
\label{figure:beamstructure}
\end{figure}

Figure \ref{figure:beamstructure} shows some examples of how differential systematics manifest in the beam pattern. The left panel shows the differential ellipticity systematic results in the differenced beam having a quadrupole feature which effectively leaks the second spatial derivative of the temperature signal into the polarization \citep{PhysRevD.77.083003}. At power spectrum level this results in the leakage to the $B$-mode going according to an $\sim \ell^4 C^{TT}_{\ell}$ term, with the $E$-mode also experiencing this along with a $\sim \ell^2 C^{TE}_{\ell}$ effect due to a cross term between the systematic and on sky signal \citep{2020arXiv200800011M}.

The middle panel shows the differential pointing is a dipole effect that effectively leaks the first spatial derivative of the temperature signal into the polarization. At power spectrum level this results in the leakage to the polarization being dominated by a $\sim \ell^2 C^{TT}_{\ell}$ term \citep{Wallisetal2016,2020arXiv200800011M}.

The right panel shows that the differential gain is a monopole effect which results in direct leakage of the temperature to polarization. At power spectrum level this results in the temperature leaking into polarization directly according to a $\sim C^{TT}_{\ell}$ term \citep{Wallisetal2016,2020arXiv200800011M}.


\bsp	
\label{lastpage}
\end{document}